\documentclass[a4paper, 12pt]{article}

\usepackage[utf8]{inputenc}      
\usepackage{amsmath, amssymb}    
\usepackage{graphicx}          
\usepackage{booktabs}                          
\usepackage{hyperref}            
\usepackage{geometry}             
\usepackage{setspace}
\usepackage[table]{xcolor}
\usepackage{booktabs}
\usepackage[table]{xcolor}  
\definecolor{rowgray}{gray}{0.90} 
\usepackage{caption}
\usepackage{float}
\usepackage{enumitem}
\usepackage[authoryear]{natbib}
\usepackage{multirow}
\setcitestyle{aysep={}}

\title{Racial Comparability in Authoritarianism Scales:\\
Latent Beliefs or Biased Measurement?}

\author{
Bang Quan Zheng \\\\
University of Texas at Austin \\
\texttt{bangquan@ucla.edu}
}

\date{2026}

\begin{document}

\maketitle

\begin{singlespace}
\begin{abstract}

\noindent Racial differences in authoritarianism are widely used to explain variation in political attitudes, yet it is unclear whether they reflect true latent differences or measurement artifacts. Using anchor-based multi-group confirmatory factor analysis across multiple nationally representative surveys, this paper examines measurement equivalence in the standard child-rearing authoritarianism battery. We find systematic differences in how respondents use response categories across groups Accounting for this non-invariance alters—but does not eliminate—racial differences in authoritarianism; African Americans continue to exhibit higher latent authoritarianism under partial scalar invariance. However, conventional multi-item scales substantially attenuate the association between authoritarianism and policy attitudes. These results show that measurement non-invariance is not merely a technical concern but can meaningfully shape substantive inferences about racial differences and their political consequences, underscoring the importance of explicit measurement modeling in studies of public opinion and political behavior.

\vspace{4em}
\end{abstract}
\end{singlespace}

\begin{center}
[Word Count: 7,860]
\end{center}

\begin{center}
\textbf{}
\end{center}

\vspace{0.25in}

\begin{center}
\textbf{\textit{Keywords:}} Authoritarianism, measurement invariance, DIF, SEM
\end{center}

\doublespacing
\newpage

\section{Introduction}

Why do widely used measures of authoritarianism produce racial differences that appear inconsistent with observed political behavior? Standard accounts define authoritarianism as a predisposition toward conformity, punitive social control, and support for exclusionary policies \citep{engelhardt2021, Feldman1997, Hethering2011, Stenner2005, Federico2026}. Yet a persistent empirical puzzle remains: Black Americans score higher than White Americans on common authoritarianism scales, while simultaneously expressing more egalitarian policy preferences and stronger alignment with the Democratic Party. 

This puzzle has direct implications for how authoritarianism is understood in political psychology. One possibility is that the inconsistency does not reflect differences in underlying dispositions, but rather differences in how survey items function across groups—raising concerns about differential item functioning (DIF) and, more broadly, measurement non-invariance in widely used scales.  Existing explanations fall into two broad camps. One emphasizes heterogeneous social meaning and socialization, suggesting that authoritarianism may be interpreted differently across groups due to divergent historical and social experiences, including variation in intragroup norms and identity-based concerns \citep{Harris2012, Cohen1999, Jefferson2023, ParkerTowler2019}. The other locates the discrepancy in measurement, arguing that standard instruments may fail to capture equivalent latent constructs across racial groups \citep{RN925, Pietryka2022}. We bridge these accounts by demonstrating that measurement non-invariance is not merely a statistical artifact, but a quantifiable manifestation of how social and political contexts reshape the meaning of survey items for different groups.

This paper takes up the measurement account directly and shows that estimates of the authoritarianism gap are sensitive to measurement non-invariance. The widely used four-item scale assumes that respondents interpret items and response categories equivalently across groups—an assumption that is rarely tested. We find systematic, item-specific threshold non-invariance, most pronounced in the child-rearing battery, alongside a subset of stable items that provide the basis for identification under partial invariance. Importantly, the direction and magnitude of these differences vary across survey contexts, indicating that DIF is not fixed but contingent on question wording, survey design, and broader informational environments.

We formalize these concerns using an anchor-based multi-group confirmatory factor analysis (MG-CFA) framework for ordinal data that explicitly models differential item functioning. This approach separates latent trait differences from group-specific response tendencies while preserving identification of group means. We apply this framework across multiple nationally representative datasets, including the 2008 and 2016 American National Election Studies (ANES), the 2008 Cooperative Campaign Analysis Project (CCAP), and the 2016 Nationscape survey (NSCP). We find that accounting for measurement non-invariance reduces—but does not eliminate—the Black–White gap in authoritarianism: African Americans still score higher on the latent trait. At the same time, correcting the measurement model strengthens the relationship between authoritarianism and a range of policy attitudes, suggesting that conventional multi-item scale attenuates substantively meaningful associations.

This paper makes three contributions. First, it develops and applies an anchor-based MG-CFA framework for ordinal survey data that explicitly models DIF while preserving identification of latent group differences, providing a flexible strategy for assessing cross-group measurement equivalence in political psychology scales. Second, it shows that widely used authoritarianism measures exhibit systematic threshold non-invariance across racial groups, indicating that respondents do not use response categories in strictly comparable ways; as a result, estimates of the Black–White authoritarianism gap are sensitive to measurement assumptions and conventional approaches may partially conflate measurement artifacts with latent differences. Third, it demonstrates that failing to account for measurement non-invariance can meaningfully distort substantive inferences about authoritarianism’s political consequences, as scale-based estimates tend to attenuate associations between authoritarianism and policy attitudes, leading to understated conclusions about its behavioral relevance.

\section{Authoritarianism: Concept \& Measurement}

Debates over authoritarianism concern whether it reflects a stable personality trait or a socially shaped attitudinal orientation. Classic work conceptualizes authoritarianism as a relatively enduring predisposition toward order and deference to authority \citep{RN1066, Feldman1997, RN158}, while broader theories of public opinion emphasize the role of informational environments and contextual cues in shaping expressed attitudes \citep{Zaller1992, ZallerFeldman1992}.

Although authoritarianism is typically conceptualized as a stable predisposition toward conformity and social control, growing evidence suggests that its empirical expression is shaped by social and political context. This raises a central measurement concern: commonly used scales may not function equivalently across racial groups, potentially conflating latent predispositions with group-specific interpretations of survey items. 

The child-rearing items used to measure authoritarianism are particularly susceptible to such variation because their meanings are socially and culturally contingent. Preferences for children who “respect elders” may reflect norms of intergenerational respect rather than orientations toward hierarchy, while valuing “obedience” may reflect adaptive responses in contexts characterized by social vulnerability or institutional distrust rather than ideological support for social control \citep{RN925, Hethering2011, Federico2026}. Similarly, “independence” may signal individual autonomy in some contexts but conflict with norms emphasizing family obligation in others. These interpretive differences can shape how respondents map latent attitudes onto response categories, potentially generating threshold variation that does not reflect differences in the underlying trait. This raises a central question: do conventional child-rearing batteries measure the same construct across racial groups?

\subsection{Cross-Group Latent Mean Identification}

A central challenge in the study of authoritarianism and related political attitudes is determining whether observed group differences reflect genuine variation in the latent trait or artifacts of measurement non-invariance. Rather than reflecting random error or simple "noise," survey responses are filtered through group-specific social and historical lenses that shape how items are understood and mapped onto response scales \citep{King2004}. In the context of racial politics, this suggests that the "authoritarianism paradox" may stem from a systematic divergence in item interpretation: what appears to be a higher level of authoritarianism may instead reflect a distinct cultural valuation of the survey’s indicators. This perspective aligns with the broader argument that substantive inference in social science is inherently model-dependent and highly sensitive to the validity of measurement assumptions. Research on racial differences in child-rearing scales has shown that multi-item batteries—especially those with ordinal indicators—are vulnerable to DIF across social groups. These studies correctly emphasize that comparisons based on composite scale scores may conflate substantive attitudinal differences with heterogeneous response processes. Yet diagnosing non-invariance is only the first step; recovering unbiased latent mean differences requires additional identification constraints, which are often left unspecified. 

Without at least one threshold- and loading-invariant indicator, group-specific latent means cannot be separated from item intercept shifts. When no such anchors are imposed, it is impossible to distinguish true differences in the underlying latent trait from item-specific bias. This identification problem is especially acute in short batteries, such as the four-item child-rearing scale, where each item heavily influences the latent construct. Although standard invariance testing procedures normalize latent means to zero during model estimation \citep{Muthen2007}, this does not resolve identification once scalar or threshold noninvariance is detected. Partial invariance structures, anchored by substantively defensible items, are therefore essential for empirically recoverable latent mean differences.

A parallel line of research has documented that many widely used ANES scales fail basic psychometric diagnostics and often conflate multiple latent dimensions \citep{Pietryka2022}. While these studies highlight measurement limitations in legacy survey instruments, their primary contribution has been diagnostic: identifying dimensional instability and item misfit rather than providing methods for valid cross-group comparisons under noninvariance. This study addresses this gap by demonstrating how partial scalar invariance with anchor items can correct item-level bias and recover unbiased latent group estimates. 

The ordinal nature of the child-rearing items introduces additional measurement considerations. In ordered categorical models, observed responses depend on both factor loadings and item-specific thresholds that map the latent trait onto discrete response categories. When thresholds vary across groups, individuals with equivalent latent authoritarianism may systematically differ in their response category usage. Consequently, observed differences in scale distributions may reflect differences in threshold parameters rather than differences in the underlying trait. Without explicit threshold constraints or anchor items, cross-group comparisons risk conflating latent differences with response-scale artifacts.

These measurement features also have downstream implications for substantive inference. When item thresholds and measurement precision vary across groups, composite scores contain unequal measurement error. This can attenuate estimated correlations with external political attitudes, potentially producing weaker observed associations in one group relative to another. Apparent differences in the relationship between authoritarianism and policy preferences may therefore reflect differences in measurement precision rather than substantively weaker attitudinal linkages.

Existing research documents substantial uneven item functioning in common authoritarianism measures but leaves unresolved the identification challenges required for valid cross-group latent mean comparisons. In particular, there is no widely used framework that simultaneously addresses item-level noninvariance and the identification constraints of short ordinal batteries. Building on these diagnostic insights, this study develops an anchor-based approach that combines invariant items with explicit modeling of item-specific thresholds and differential item functioning. This framework improves separation of latent attitudinal differences from measurement artifacts induced by noninvariance in the child-rearing authoritarianism scale.

\section{Research Strategy}

Because claims about racial differences in authoritarianism hinge on valid cross-group comparisons, this study prioritizes establishing measurement equivalence before drawing substantive conclusions. We implement an anchor-based MG-CFA to evaluate whether commonly used child-rearing items measure the same latent construct among Black and White respondents. 

The research strategy proceeds in three steps. First, we identify anchor items that satisfy cross-group invariance and provide a stable basis for latent mean comparisons. Second, we estimate group-specific latent means under a partial invariance framework, correcting for item-level bias in threshold and loading parameters. Third, we embed the validated measurement model within a structural equation framework to assess whether authoritarianism predicts policy attitudes consistently across groups. 

All models are estimated in \texttt{R} using the \texttt{lavaan} package \citep{rosseel2012}, with estimators appropriate for ordinal indicators. This approach ensures that observed group differences reflect substantive variation in latent authoritarianism rather than artifacts of DIF or measurement noninvariance.

\subsection{Step 1: Establishing Measurement Equivalence}

\begin{figure}[H]
    \centering
    \caption{MG-CFA of Authoritarianism (Black–White)}
    \includegraphics[width=0.7\textwidth]{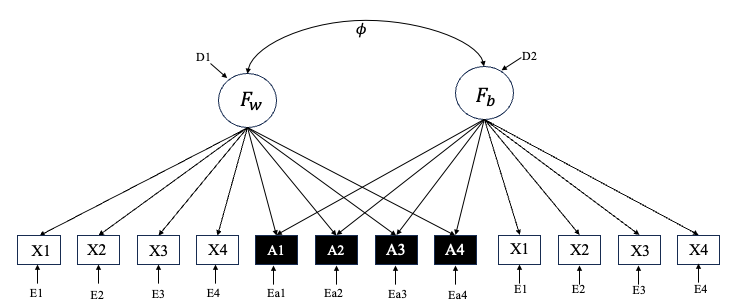} 
    \label{fig:Figure-1}
\end{figure}

Figure~\ref{fig:Figure-1} illustrates the multiple-group framework applied to Black and White respondents. Let $X_i$ denote the child-rearing indicators and $\eta_i$ the latent authoritarianism trait. The goal of this step is to determine whether at least some indicators function equivalently across groups, thereby allowing separation of true latent differences from item-specific response differences.

We begin by estimating a configural model in which all parameters are freely estimated across groups. We then evaluate metric (loading) and scalar (threshold) invariance using nested model comparisons appropriate for categorical indicators. Items that do not exhibit meaningful cross-group differences in loadings and thresholds are freed, while those that do are constrained to equality and serve as anchors. This partial invariance specification provides the basis for identifying latent means without imposing full measurement equivalence across groups. 

Because the child-rearing battery contains only four items, anchor selection is consequential: each indicator plays a substantial role in defining the latent construct. Establishing at least one invariant indicator ensures that estimated group differences reflect variation in the latent trait rather than shifts in item interpretation.

\subsection{Step 2: Latent Mean Estimation}

Authoritarianism is modeled as a latent variable within a multiple-group SEM framework. For individual $i$, let $\mathbf{y}_i = (y_{i1}, \dots, y_{iJ})'$ denote the observed binary indicators. The measurement model is:

\begin{equation}
\mathbf{y}_i = \boldsymbol{\tau}_g + \boldsymbol{\Lambda}_g \eta_i + \boldsymbol{\varepsilon}_i,
\end{equation}

\noindent where $\boldsymbol{\tau}_g$ represents group-specific thresholds, $\boldsymbol{\Lambda}_g$ factor loadings, and $g \in \{\text{White}, \text{Black}\}$ indexes racial group membership.

With categorical indicators, observed responses are linked to the
latent trait through thresholds rather than intercepts
\citep{Muthen1984, RN1030}. Cross-group differences in thresholds
can therefore generate observed mean differences even when latent
trait levels are identical, a form of DIF
that biases naive comparisons \citep{Davidov2014}. By constraining only
anchor-item loadings and thresholds to equality across groups, latent
mean differences become separately identified from item-level shifts
\citep{Steenkamp1998, Vandenberg2000}. Under this structure, estimated
group differences can be interpreted as substantive differences in
authoritarian predispositions rather than artifacts of DIF.

\subsection{Step 3: Structural Relationships with Policy Attitudes}

In the final step, the validated measurement model is embedded within a structural equation framework to examine whether authoritarianism predicts policy attitudes similarly across racial groups. Let $Y_i$ denote a policy attitude outcome. The structural model is:

\begin{equation}
Y_i = \alpha_g + \beta_g \eta_i + \mathbf{\gamma}_g \mathbf{X}_i + \zeta_i,
\end{equation}

\noindent where $\mathbf{X}_i$ includes party identification, income, age, education, gender, and homeownership status. This specification allows the effect of authoritarianism ($\beta_g$) to vary across groups, enabling direct tests of whether the political consequences of authoritarian predispositions differ between Black and White respondents.

Estimation jointly models the measurement and structural components, integrating over the distribution of the latent factor rather than relying on factor score estimates. This ensures that both measurement noninvariance and estimation uncertainty are fully propagated into estimates of policy relationships. This approach propagates measurement uncertainty into the structural analysis and ensures that conclusions about racial differences in authoritarianism and its political effects are not driven by item-level measurement bias.

\section{Factor Invariance with Categorical Data}

Survey items often use discrete response categories rather than continuous scales. Differences in how respondents use these categories can create the illusion of trait differences even when the underlying psychological construct is similar. This issue is particularly salient for the four-item child-rearing scale. To address it, we employ MG-CFA with categorical indicators, which separates genuine latent trait differences from group-specific response tendencies.

\subsection{Latent Response Model for Categorical Items}

Each observed categorical response is assumed to arise from an underlying continuous latent variable. Let $Y_j^*$ denote the latent response associated with item $j$. Threshold parameters $\tau_{jk}$ define the cut points that map $Y_j^*$ onto observed response categories $1, 2, \dots, K$.

Stacking the latent responses across items yields the vector $\mathbf{Y}^*$, which is modeled as

\begin{equation}
\mathbf{Y}^* = \mathbf{v} + \boldsymbol{\Lambda}\boldsymbol{\xi} + \boldsymbol{\epsilon},
\end{equation}

\noindent where $\boldsymbol{\xi}$ denotes the latent authoritarianism factor,
$\boldsymbol{\Lambda}$ is the factor loading matrix linking items to the latent
trait, $\mathbf{v}$ is a vector of latent intercepts, and $\boldsymbol{\epsilon}$
represents item-specific residuals. The thresholds $\tau_{jk}$ translate the
continuous latent responses into observed ordinal categories.

\subsection{Anchor-Based Framework of Partial Scalar Invariance}

Anchor-based partial scalar invariance provides a flexible framework for assessing measurement comparability when full invariance does not hold. In MG-CFA, measurement invariance is evaluated by testing whether factor loadings and item thresholds are equivalent across groups; constraining loadings corresponds to metric invariance, while constraining thresholds corresponds to scalar invariance.

In applied survey research, full scalar invariance is often unrealistic, particularly for multi-item scales that combine substantively heterogeneous indicators and are therefore more susceptible to DIF across groups. When full invariance fails, partial scalar invariance can be achieved by freeing non-invariant item parameters while constraining a subset of anchor items. This strategy allows the model to recover a common latent scale despite the presence of DIF in loadings and thresholds.

We additionally estimate a no-anchor partial invariance specification, which imposes full scalar invariance across groups, as a diagnostic benchmark for assessing the consequences of mis-specified measurement invariance assumptions.

Partial scalar invariance is sufficient for identifying latent mean differences \citep{Muthen1989}, with the reference group (White respondents) normalized to zero. Under this specification, equality constraints on anchor items define a common scale, ensuring that cross-group differences are attributed to the latent trait rather than measurement noninvariance.

\subsection{Monte Carlo Assessment of Measurement Bias}

To clarify both the identification logic and the finite-sample performance of the proposed approach, we proceed in two steps. First, we describe how partial scalar invariance with anchor items identifies latent mean differences in the presence of DIF. Second, we use Monte Carlo simulations to evaluate the performance of this strategy under varying degrees of measurement non-invariance.

To illustrate the consequences of measurement non-invariance, we conducted a Monte Carlo simulation based on a canonical four-anchor, four-DIF item structure commonly used in authoritarianism batteries. The simulation evaluates how anchor-based partial scalar invariance performs relative to conventional scale aggregation when DIF is present. In each replication, we generated a sample of $N = 1{,}000$ respondents, evenly divided between White and Black groups. The latent authoritarianism trait for respondent $i$ in group $g$ was drawn from:

\[
\xi_i \sim N(\mu_g, 1),
\]

\noindent where $\mu_g$ denotes the group-specific latent mean. The true latent difference was set to $\mu_{\text{Black}} - \mu_{\text{White}} = 0.2$. Each respondent was assigned eight ordinal indicators generated according to the latent response model:

\[
Y_{ij}^* = \lambda_j \xi_i + \epsilon_{ij},
\]

\noindent where $\lambda_j$ is the item loading and $\epsilon_{ij} \sim N(0,1)$ is an item-specific disturbance term. Continuous responses were discretized into four ordered categories using fixed thresholds.

For each simulated dataset, we estimate group differences using three approaches: (1) a conventional multi-item scale based on mean item scores, (2) a full scalar invariance model that constrains all item parameters to equality across groups, and (3) an anchor-based partial scalar invariance model identified using the four invariant items. The anchor-based specification establishes a common latent metric across groups, allowing latent means to be identified net of item-specific threshold variation. In contrast, the conventional scale and full invariance model both conflate latent differences with unmodeled threshold shifts under conditions of non-invariance, yielding biased estimates of the group difference.

To introduce measurement non-invariance, we simulate eight ordinal items, of which four are designated as DIF items and four as invariant anchors. For DIF items, thresholds for Black respondents are shifted by an amount $\delta$, while anchor items are held invariant by construction:

\[
\tau_{jk}^{\text{B}} = \tau_{jk}^{\text{W}} + \delta.
\]

The anchor-based partial invariance model correctly reflects the data-generating process by constraining only the invariant (anchor) items, whereas the full scalar invariance model is misspecified under conditions of DIF because it imposes equality on all items. The conventional multi-item scale is constructed using only the four DIF items, mirroring standard practice in applications that do not account for measurement non-invariance in the child-rearing battery. This setup enables a direct comparison of estimator performance under a known data-generating process with systematic threshold variation across groups.

The magnitude of $\delta$ was systematically varied across simulation conditions ($\delta = 0$ to $0.5$) to represent increasing levels of measurement non-invariance. For each simulated dataset, group differences were estimated using (1) a conventional multi-item scale based on mean item scores, and (2) latent mean differences under partial scalar invariance, identified using four pre-specified anchor items generated to be invariant by construction. \\

\begin{figure}[htbp]
    \centering
    \caption{Evaluating Measurement Bias in Simulated Scales}
    \includegraphics[width=0.95\textwidth]{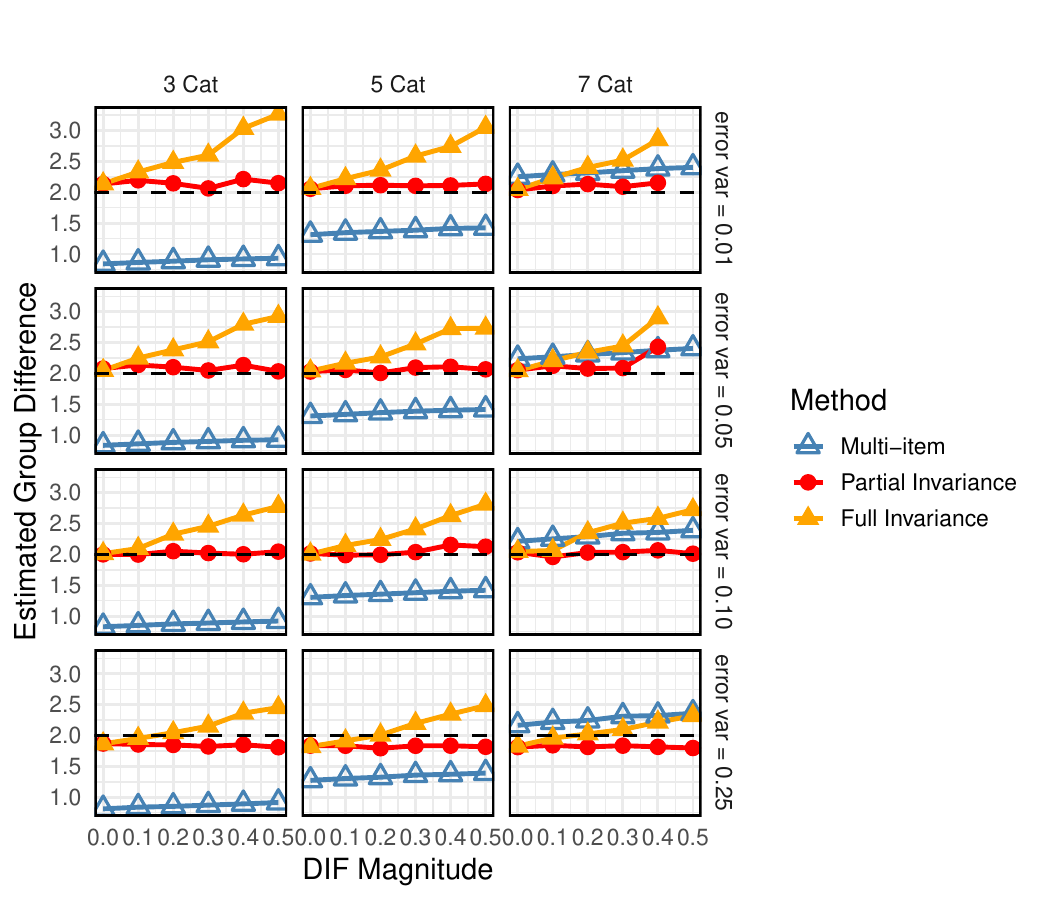}
    \label{fig:DIF_simulation}
\end{figure}

To systematically evaluate performance, we vary factor loadings and residual variances across measurement conditions and combine these with four levels of threshold shift ($\delta$), producing the panel structure in Figure~\ref{fig:DIF_simulation}. Each condition is replicated 500 times, and we report the average estimated group difference relative to the true latent difference of 0.2.

Figure~\ref{fig:DIF_simulation} summarizes the performance of multi-item aggregation, full scalar invariance, and partial scalar invariance across these conditions. Several consistent patterns emerge. First, partial scalar invariance models most accurately recover the true group difference across nearly all conditions, particularly as $\delta$ increases. By allowing selected item thresholds to vary across groups while constraining anchor items, this approach effectively mitigates bias induced by DIF. In contrast, conventional multi-item aggregation exhibits substantial bias under moderate and high DIF, reflecting its inability to accommodate item-specific group shifts.

Second, full scalar invariance performs adequately when measurement error is low and response categories are sufficiently granular, but its performance deteriorates as measurement noise increases. In these settings, strict equality constraints appear to amplify estimation instability, especially when item responses are noisy.

Third, measurement precision—captured by both the number of response categories and the magnitude of residual variance—systematically shapes estimator performance. Fewer response categories increase discretization error and reduce information about the latent trait, leading to greater bias in conventional aggregation approaches, while finer scales (e.g., 7-category items) improve accuracy across all methods.

Across conditions, as $\delta$ increases (left to right across panels), conventional scales increasingly deviate from the true value due to unmodeled threshold shifts. By contrast, partial scalar invariance remains robust, consistently recovering the true latent difference even under substantial DIF and moderate measurement noise.

Finally, the appendix (Figures~\ref{app:sim-1}--\ref{app:sim-3}) reports supplementary simulations with alternative factor loading specifications, including settings where both loadings and thresholds are constrained to equality across groups. These additional analyses confirm the robustness of the main findings under stronger measurement assumptions.

\section{Data and Measurement}

In this study, we will utilize the 2008 and 2016 ANES data, the 2008 Cooperative Campaign Analysis Project (CCAP) data, and the 2016 Nationscape (NSCP) data to examine the psychological construct of child-rearing scales illustrated in Figure 1. These datasets are nationally representative and were collected through probability sampling, making them well suited for generalization and cross-validation purposes.

\subsection{Anchor Item Justification}

Anchor items provide a common metric for cross-group comparisons in MG-CFA by fixing the latent scale. We identify a subset of items that exhibit stable factor loadings across groups, based on both theoretical considerations and empirical tests of invariance. These anchors are constrained to equality, while all remaining item parameters are allowed to vary, enabling identification under partial scalar invariance without imposing full cross-group equivalence.

To enhance robustness, anchor items are drawn from conceptually distinct domains across datasets. This reduces dependence on any single item set and ensures that identified non-invariance in the child-rearing battery reflects item-specific differences rather than artifacts of scale construction. Table~\ref{tab:anchor_items_summary} summarizes the anchor items used in each dataset; full wording and survey codes are provided in the Appendix.

To improve robustness, anchor items are drawn from conceptually distinct domains across datasets. This reduces dependence on any single item set and helps ensure that identified non-invariance in the child-rearing battery reflects item-specific differences rather than artifacts of scale construction.

\begin{table}[h!]
\centering
\caption{Summary of Anchor Items Across Datasets}
\label{tab:anchor_items_summary}
\begin{tabular}{p{3cm} p{6cm} p{6cm}}
\toprule
\textbf{Dataset} & \textbf{Anchor Items} & \textbf{Substantive Rationale} \\
\midrule
2008 ANES & Four cognitive/personality items: problem complexity, responsibility for thinking, enjoyment of complex thinking, beliefs about personal change & Conceptually distinct from authoritarianism; widely understandable; avoids culturally specific meanings \\ 
\midrule
2008 CCAP & Four optimism/pessimism items from LOT-R: relaxation, enjoyment of friends, staying busy, emotional stability & Neutral affective orientation; unrelated to authoritarianism; widely interpretable \\ 
\midrule
2016 ANES & Four personality/cognition items: “R has many more opinions than average,” “R likes to have strong opinions even when not personally involved,” “Sympathetic, warm” self-description, “Conventional, uncreative” self-description & Conceptually unrelated to authoritarianism; stable reference points for partial scalar identification; minimal correlation among items \\ 
\midrule
2016 NSCP & Four feeling-thermometer items toward Latino, Asian, Jewish, and alt-right groups & Conceptually unrelated to authoritarianism; stable reference points for partial scalar identification \\ 
\bottomrule
\end{tabular}
\begin{flushleft}
\footnotesize \textit{Note:} LOT-R = Life Orientation Test. \\
\footnotesize \textit{} Full item wording and survey codes for each dataset are provided in the Appendix.
\end{flushleft}
\end{table}

Statistical validation complements this substantive rationale. Anchor items are selected based on both theoretical considerations and empirical evidence of measurement invariance, including minimal differential item functioning and stable factor loadings across groups. Items failing these criteria are freed under the partial scalar invariance specification.

In practice, anchor selection is context-dependent. In cases where no clearly neutral items are available, we draw anchors from conceptually distinct domains that are not intended to measure authoritarianism directly but demonstrate empirical invariance across groups. Consistent with prior work, low inter-item correlations are acceptable given that anchors are not required to form a coherent latent scale \citep{Vandenberg2000, Steenkamp1998, Davidov2014}.

\subsection{Anchored Measurement Invariance Test}

To address potential measurement issues, we select four anchor items whose functioning is expected to be comparable across groups. MG-CFA results (see Appendix) show that these items exhibit similar variance patterns across groups, supporting their use as anchors. Tables A1–A4 present the corresponding correlation matrices.

The child-rearing items are dichotomous (0/1) and were modeled as ordered categorical indicators. The anchor items, by contrast, use ordered response categories and were modeled accordingly. Because both sets of items are treated as categorical indicators within the same MG-CFA framework, the anchor-based identification strategy remains appropriate and does not introduce scale heterogeneity. All models were estimated using weighted least squares mean and variance (WLSMV) estimator, which is designed for models with categorical indicators. Measurement invariance is evaluated using likelihood ratio tests (LRTs), which compare the fit of nested models with increasingly restrictive equality constraints across groups. 

Measurement invariance for categorical indicators does not require thresholds to be held equal across groups for the purposes of latent mean identification when items are used as anchors. Anchor items establish a common metric for the latent variable through equality constraints on factor loadings, rather than on thresholds \citep{RN1030}. As \citet{KimYoon2011} show, anchor-based identification in multi-group CFA and IRT requires metric invariance of the anchor items, while allowing thresholds (or item intercepts) to vary freely without biasing latent mean comparisons. Similarly, \citet{Muthen2012} demonstrates that for ordinal indicators, fixing a subset of loadings across groups is sufficient for scale identification, whereas full threshold invariance is neither required nor generally realistic. Accordingly, constraining anchor loadings while permitting threshold non-invariance is both theoretically justified and standard practice in contemporary anchor-item approaches.

\begin{table}[htbp]
\centering
\caption{Anchor Item Validation}
\label{tab:lrt_anchor}

\begin{minipage}{0.48\textwidth}
\centering
\small
\begin{tabular}{lcccccc}
\hline
 & $df$ & $\chi^2$ & $\Delta\chi^2$ & $\Delta df$ & $p$-value \\
\hline
\multicolumn{6}{c}{\textbf{2008 ANES}} \\
Configural & 4  & 10.259 &        &     &        \\
Metric     & 7  & 14.433 & 4.077  & 3   & 0.253  \\
Scalar     & 12 & 51.983 & 38.500 & 5   & 0.001 \\
\hline
\multicolumn{6}{c}{\textbf{2008 CCAP}} \\
Configural & 4 & 301.52 &       &     &   \\
Metric   & 7 & 305.30 & 3.5229 & 3   & 0.318 \\
Scalar    & 10 & 318.31 & 24.961 & 3   & 0.000 \\

\hline
\end{tabular}
\end{minipage}
\hfill
\begin{minipage}{0.48\textwidth}
\centering
\small
\begin{tabular}{lcccccc}
\hline
 & $df$ & $\chi^2$ & $\Delta\chi^2$ & $\Delta df$ & $p$-value \\
\hline
\multicolumn{6}{c}{\textbf{2016 ANES}} \\
Configural & 4 & 6.925 &       &     &   \\
Metric    & 7 & 9.039 & 2.658 & 3   & 0.448 \\
Scalar    & 10 & 11.151 & 4.106 & 3   & 0.250 \\
\hline
\multicolumn{6}{c}{\textbf{2016 NSCP}} \\
Configural & 4  & 38.232 &       &     &   \\
Metric     & 7  & 43.409 & 5.177 & 3   & 0.217 \\
Scalar     & 14 & 65.525 & 22.116 & 7  & 0.000 \\
\hline

\end{tabular}
\end{minipage}

\end{table}

Table~\ref{tab:lrt_anchor} reports likelihood-ratio tests of measurement invariance between Black and White respondents across four datasets: 2008 ANES, 2008 CCAP, 2016 ANES, and 2016 NSCP. For each dataset, the table presents $\chi^2$ statistics for the sequential evaluation of configural, metric, and scalar invariance. Across datasets, scalar invariance is not supported, indicating systematic differences in response thresholds. In contrast, metric invariance generally holds, suggesting that item loadings are comparable across groups and that the underlying factor structure is stable. These results justify a partial invariance specification in which loading equality is maintained while threshold parameters are allowed to vary.

\subsection{Statistical Analysis}
We assess measurement invariance of the child-rearing authoritarianism scale using a sequence of increasingly restrictive MG-CFA models. Model 1 (configural) allows all parameters to vary freely, establishing that the factor structure is comparable across groups. Model 2 (metric) constrains factor loadings to equality; model fit does not meaningfully deteriorate, consistent with metric invariance and indicating that items load similarly on the latent construct across groups.

Model 3 (scalar) additionally constrains item thresholds to equality. Across datasets, this specification produces significant misfit, indicating systematic differences in category usage across racial groups. Model 4 further constrains latent means, yielding additional misfit, consistent with the presence of threshold non-equivalence. Together, these results suggest that observed group differences are driven primarily by response-scale heterogeneity rather than differences in the measurement structure of the latent trait. \\

\begin{table}[htbp]
\label{tab:table2}
\centering
\scriptsize

% ----------------------------
% General Title for Both Tables
% ----------------------------
\caption{Measurement Invariance Across Studies}
\label{tab:table2}

\begin{minipage}{0.48\textwidth}
\centering

\textbf{2008 ANES} \\[2mm]  % <-- source above table

\begin{tabular}{rlccccc}
\hline
 & Model & $df$ & $\chi^2$ & $\Delta \chi^2$ & $\Delta df$ & $p$-value \\
\hline
1 & \textbf{Configural}     & 40 & 348.98 &        &     &        \\
2 & \textbf{Metric}         & 47 & 354.37 &  4.554 & 7   & 0.714  \\
3 & \textbf{Scalar}         & 52 & 414.17 & 59.709 & 5   & 0.001 \\
4 & \textbf{Scalar+FVs} & 53 & 460.40 & 22.701 & 1   & 0.001 \\
\hline
\end{tabular}

\end{minipage}
\hfill
\begin{minipage}{0.48\textwidth}
\centering

\textbf{2008 CCAP} \\[2mm]  % <-- source above table

\begin{tabular}{rlccccc}
\hline
 & Model & $df$ & $\chi^2$ & $\Delta \chi^2$ & $\Delta df$ & $p$-value \\
\hline
 & \textbf{Configural}     & 40 & 2848.90 &        &     &      \\
 & \textbf{Metric}         & 47 & 2864.50 & 10.585 & 7   & 0.158 \\
 & \textbf{Scalar}         & 50 & 2894.50 & 51.483 & 3   & 0.000 \\
 & \textbf{Scalar+FVs} & 51 & 3228.80 & 243.774 & 1   & 0.000 \\
\hline
\end{tabular}

\end{minipage}

\end{table}

\begin{table}[htbp]
\centering
\scriptsize
\label{tab:mi_2016}

\begin{minipage}{0.48\textwidth}
\centering

\textbf{2016 ANES} \\[2mm]

\begin{tabular}{rlccccc}
\hline
 & Model & $df$ & $\chi^2$ & $\Delta\chi^2$ & $\Delta df$ & $p$-value \\
\hline
1 & \textbf{Configural}     & 40 & 48.739  &         &     &       \\
2 & \textbf{Metric}         & 47 & 58.859  &  7.029  & 7   & 0.426 \\
3 & \textbf{Scalar}         & 50 & 68.256  & 15.495  & 3   & 0.001 \\
4 & \textbf{Scalar+FVs} & 51 &115.657  & 30.550  & 1   & 0.000 \\
\hline
\end{tabular}

\end{minipage}
\hfill
\begin{minipage}{0.48\textwidth}
\centering

\textbf{2016 NSCP} \\[2mm]

\begin{tabular}{rlccccc}
\hline
 & Model & $df$ & $\chi^2$ & $\Delta\chi^2$ & $\Delta df$ & $p$-value \\
\hline
 & \textbf{Configural}     & 40 & 3001.7 &         &     &        \\
 & \textbf{Metric}         & 47 & 3018.8 & 12.140  & 7   & 0.096  \\
 & \textbf{Scalar}         & 56 & 3135.4 &137.470  & 9   & 0.000  \\
 & \textbf{Scalar+FVs} & 57 & 3218.0 & 41.243  & 1   & 0.000 \\
\hline
\end{tabular}

\end{minipage}

\end{table}

Overall, the results indicate that Black–White differences in authoritarianism arise mainly from how respondents map the latent trait onto item responses (thresholds), not from differences in how items relate to the latent construct (factor loadings). Applying an anchor-based approach, that is, holding a subset of thresholds stable while allowing others to vary, recovers meaningful latent mean differences, demonstrating that observed racial disparities reflect substantive differences rather than measurement artifacts. These findings highlight the importance of item-level analysis in cross-group comparisons and justify the use of partial scalar invariance to identify latent means. By establishing partial scalar invariance, we ensure that observed Black–White differences in authoritarianism reflect genuine psychological predispositions rather than artifacts of differential response styles, which is critical for interpreting racial disparities in policy preferences and political behavior.

\section{Latent Racial Differences in Authoritarianism}

We examine Black–White differences in authoritarianism at both the item and latent levels using a partial scalar invariance model, which separates variation in the latent construct from item-level measurement bias.

\subsection{Threshold Analysis}

To distinguish measurement bias from substantive group differences, we examine item-level racial differences in the authoritarianism scale. Because the child-rearing items are ordinal, observed responses reflect both respondents’ underlying latent authoritarianism and the thresholds that map this latent trait onto discrete response categories. If thresholds differ across groups, individuals with identical levels of latent authoritarianism may systematically select different response options. Apparent group differences may therefore arise from DIF rather than genuine variation in the latent construct.

We estimate group-specific thresholds and compute racial differences for each of the four child-rearing items as:

\begin{equation}
\Delta \tau_{item,k} = \tau_{item,k}^{\text{(B)}} - \tau_{item,k}^{\text{(W)}},
\end{equation}

\noindent where $\tau_{item,k}$ denotes the $k$-th threshold for a given item, and superscripts $(B)$ and $(W)$ refer to Black and White respondents, respectively. Positive values indicate that Black respondents require a higher level of the latent trait to endorse a given response category; negative values indicate the reverse. Systematic deviations from zero therefore signal DIF.

For latent mean comparisons, the White group’s latent mean is fixed to zero and the Black group’s mean is freely estimated. This parameterization ensures that estimated mean differences are evaluated net of identified threshold non-invariance, isolating the portion of the observed gap that reflects substantive differences in authoritarian beliefs rather than measurement artifacts.

\begin{figure}[H]
    \centering
    \caption{Item-Level Threshold Differences}
    
    % First row
    \begin{minipage}{0.45\textwidth}
        \includegraphics[width=\textwidth]{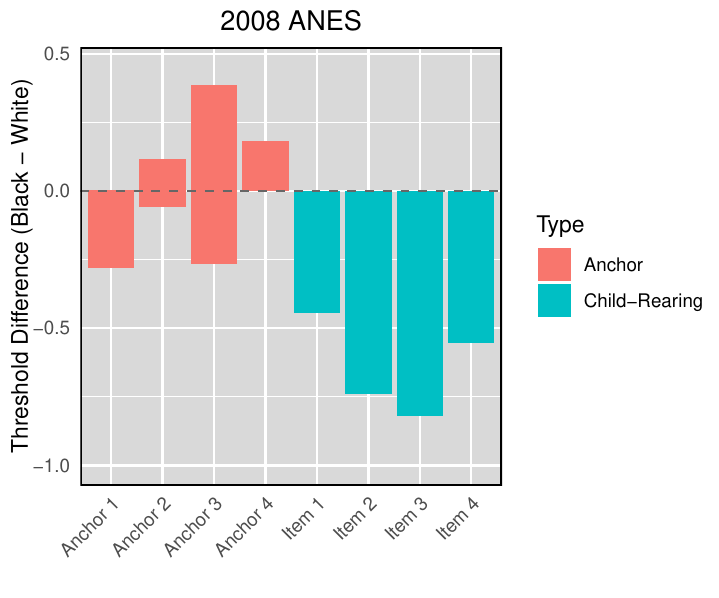}
    \end{minipage}
    \hspace{0.05\textwidth}
    \begin{minipage}{0.45\textwidth}
        \includegraphics[width=\textwidth]{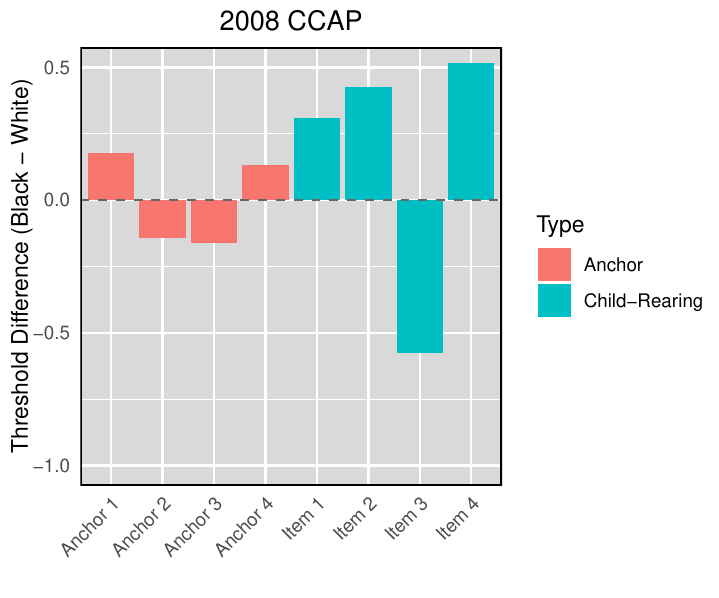}
    \end{minipage}

    % Second row
    \begin{minipage}{0.45\textwidth}
        \includegraphics[width=\textwidth]{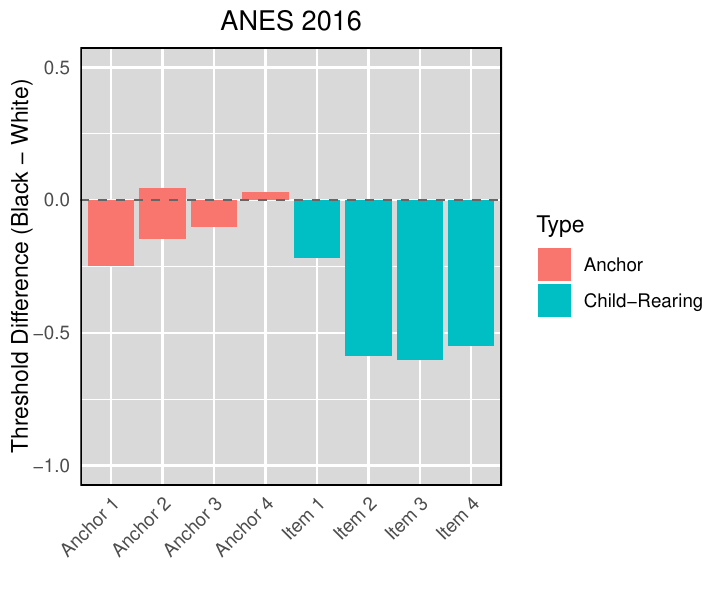}
    \end{minipage}
    \hspace{0.05\textwidth}
    \begin{minipage}{0.45\textwidth}
        \includegraphics[width=\textwidth]{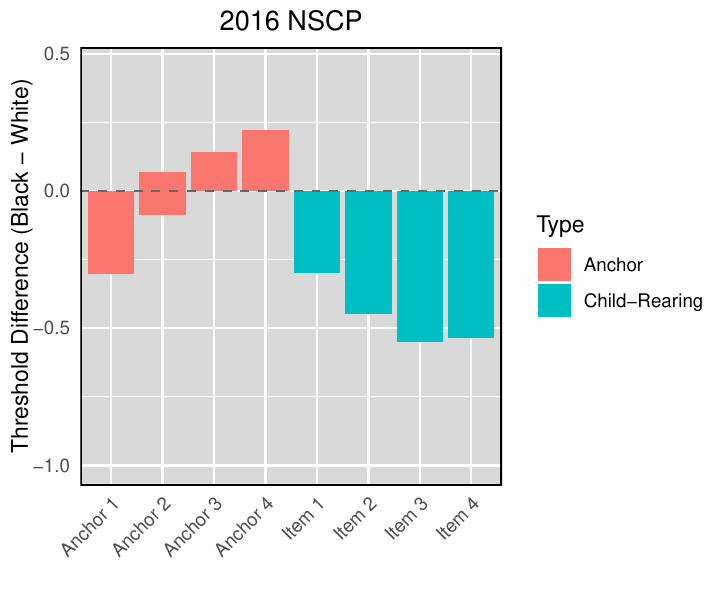}
    \end{minipage}
\caption*{\footnotesize
Note: Item 1: independence vs. respect; 
Item 2: curiosity vs. good manners; 
Item 3: self-reliance vs. obedience ; 
Item 4: considerate vs. well-behaved.
}
    \label{fig:Figure3}
\end{figure}

Figure~\ref{fig:Figure3} presents the estimated threshold differences using data from the 2008 CCAP, the 2008 and 2016 ANES, and the 2016 NSCP surveys. Large and consistent gaps—particularly for items emphasizing obedience, manners, or respect—indicate where racial comparability is weakest, whereas small or unsystematic differences suggest minimal item-level bias.

As Figure~\ref{fig:Figure3} shows, the chosen anchor items exhibit minimal differences and are balanced for Black and White respondents. For the child-rearing items, most datasets indicate that Black respondents require a lower level of latent authoritarianism to endorse a given response category compared with White respondents; in other words, Black respondents need less latent authoritarianism than Whites to select that category. Values near zero suggest that the item functions equivalently across groups, while systematic deviations signal DIF. However, the 2008 CCAP data show a different pattern: for items such as independence vs. respect, curiosity vs. good manners, and considerate vs. well-behaved, Black respondents require more latent authoritarianism than White respondents to endorse the same category. This pattern aligns with prior scholarship suggesting that the expression and measurement of authoritarianism can be contingent on context, item wording, and survey-specific factors (e.g., \citep{Feldman1997, Pietryka2022}).

\subsection{Structured Means Analysis}

Latent means for each group are estimated from the partial scalar MG-CFA model, with factor scores representing each respondent’s position on the latent authoritarianism construct net of item-level measurement differences. For identification, the latent mean for the White group is fixed to zero, so the estimated mean for the Black group directly captures the Black–White difference. Partial scalar invariance ensures that these comparisons reflect genuine differences in the underlying construct rather than DIF \citep{dimitrov2006}.

The difference in latent means is:

\begin{equation}
\Delta_{\eta} = \mu_{\eta}^{(B)} - \mu_{\eta}^{(W)},
\end{equation}

\noindent where $\mu_{\eta}^{(B)}$ and $\mu_{\eta}^{(W)}$ denote the latent means for Black and White respondents, respectively. Identification is achieved by fixing the latent mean of White respondents to zero, so that the estimated latent mean for the Black group directly captures the Black--White difference in authoritarianism. This parameterization ensures that group differences in the latent construct are estimated net of measurement non-invariance. 

Figure~\ref{fig:Figure3} compares the latent distributions of Black and White respondents under partial scalar invariance. The side-by-side presentation illustrates how correcting for measurement bias alters the inferred magnitude of racial differences. Latent means estimated under partial invariance more accurately recover underlying differences in authoritarianism between Black and White respondents. Across multiple datasets, African Americans’ latent factor scores are consistently higher than those of Whites.

\begin{figure}[H]
    \centering
    \caption{Latent Factor Scores between Blacks and Whites}
    % First row of two figures
    \begin{minipage}{0.45\textwidth}
        \centering
        \includegraphics[width=\textwidth]{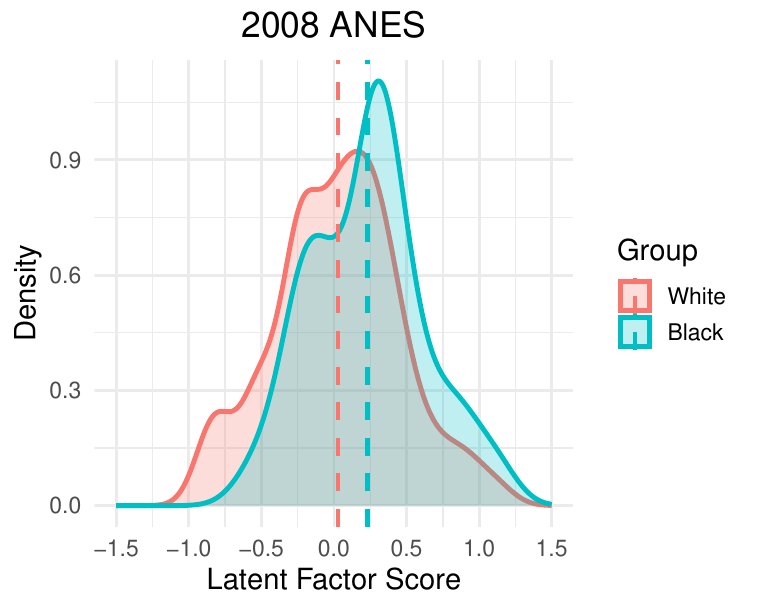}
    \end{minipage}
    \hfill
    \begin{minipage}{0.45\textwidth}
        \centering
        \includegraphics[width=\textwidth]{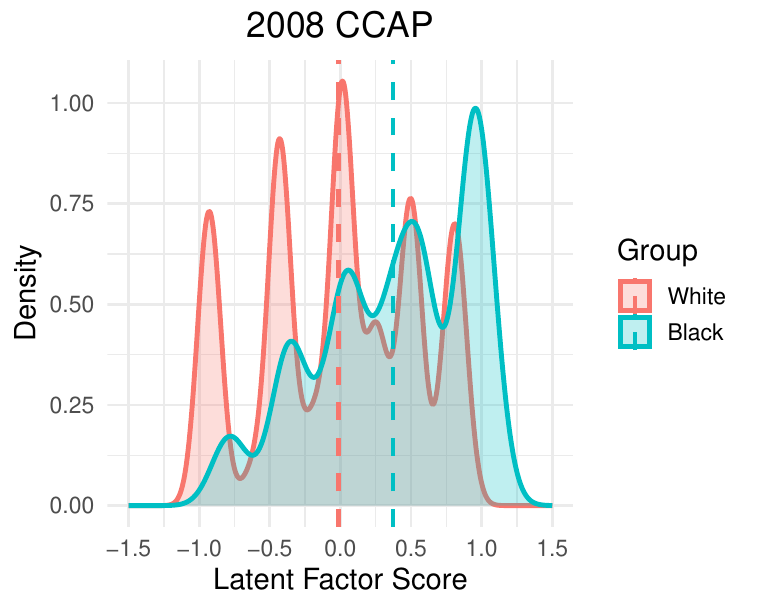}
    \end{minipage}
        \begin{minipage}{0.45\textwidth}
        \centering
        \includegraphics[width=\textwidth]{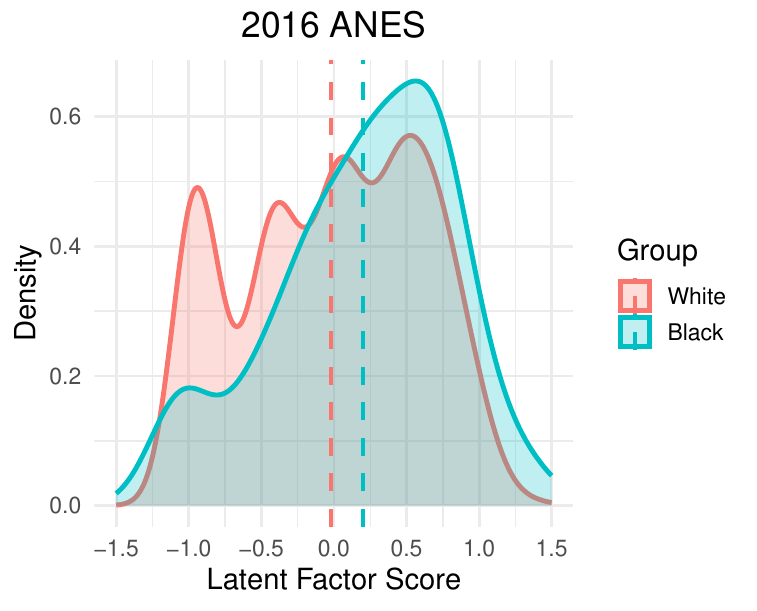}
    \end{minipage}
    \hfill
    \begin{minipage}{0.45\textwidth}
        \centering
        \includegraphics[width=\textwidth]{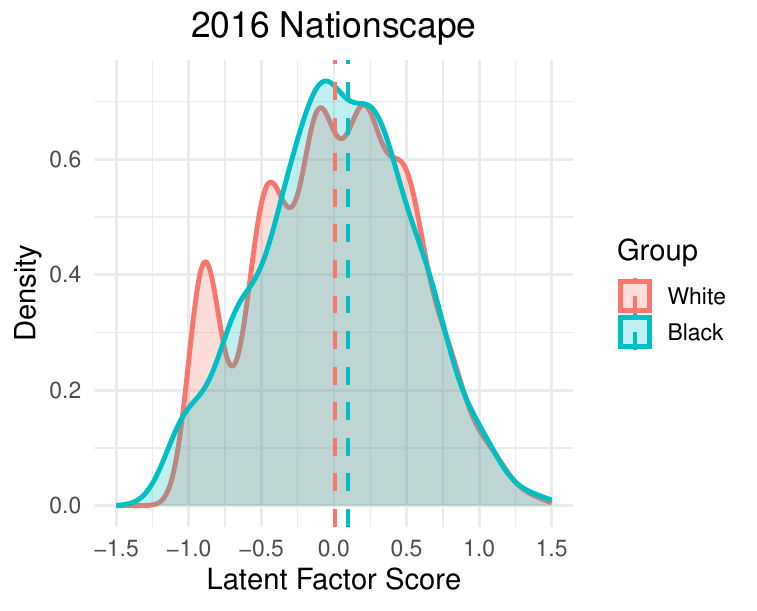}
    \end{minipage}

    \label{fig:Figure4}
\end{figure}

Taken together, the threshold and latent mean results show that although several items display modest DIF, the partial invariance model identifies meaningful latent differences in authoritarianism. After accounting for measurement noninvariance, latent mean estimates suggest higher authoritarianism among Black respondents in the ANES and Nationscape data. However, the 2008 CCAP departs from this pattern, indicating that racial differences in authoritarianism are not fully stable across survey contexts and measurement environments.

\section{Authoritarianism and Policy Preferences}

Having established Black–White differences in latent authoritarianism under partial scalar invariance, we examine the substantive implications of using bias-corrected latent scores rather than conventional multi-item scales. We estimate equivalent MG-CFA models across four datasets (2008 \& 2016 ANES, 2008 CCAP, 2016 NSCP), assessing configural, metric, and partial scalar invariance across samples. This design allows us to evaluate whether conclusions about authoritarianism’s political consequences are robust across datasets.

\subsection{Racial Gaps \& Policy in Authoritarianism}

Building on the measurement framework above, we examine how latent authoritarianism shapes Black–White differences in policy preferences. Using partial scalar MG-CFA models, we jointly estimate factor loadings, item thresholds, and residual variances, allowing a subset of thresholds to vary across groups.

Individual-level latent scores extracted from the fitted models provide bias-corrected estimates of authoritarianism for each respondent. These scores are more accurate than conventional multi-item scales, which can be attenuated by threshold heterogeneity and DIF.

We relate these latent scores to policy attitudes using a linear model:  

\begin{equation}
\text{Policy}_i = \alpha + \beta \hat{\eta}_i + \epsilon_i,
\end{equation}

\noindent where $\hat{\eta}_i$ denotes the latent authoritarianism score for respondent $i$. Because MG-CFA estimates latent means for each racial group ($\mu_g$), the implied Black–White difference in expected policy preferences is

\begin{equation}
\Delta_{\text{policy}} = \beta (\mu_{\text{Black}} - \mu_{\text{White}}),
\end{equation}

\noindent This approach addresses concerns raised in prior work that multi-item child-rearing scales may lack equivalence across Black and White respondents \citep{Pietryka2022, RN925}, potentially attenuating estimates of latent differences. The resulting latent scores provide a basis for evaluating how authoritarian predispositions map onto concrete policy preferences.

\subsection{Black-White Differences in Policy Attitudes}

\begin{figure}[H]
    \centering
    \caption{Policy Attitude Differences Across Datasets}

    % Top row
    \begin{minipage}{0.42\textwidth}
        \includegraphics[width=\textwidth]{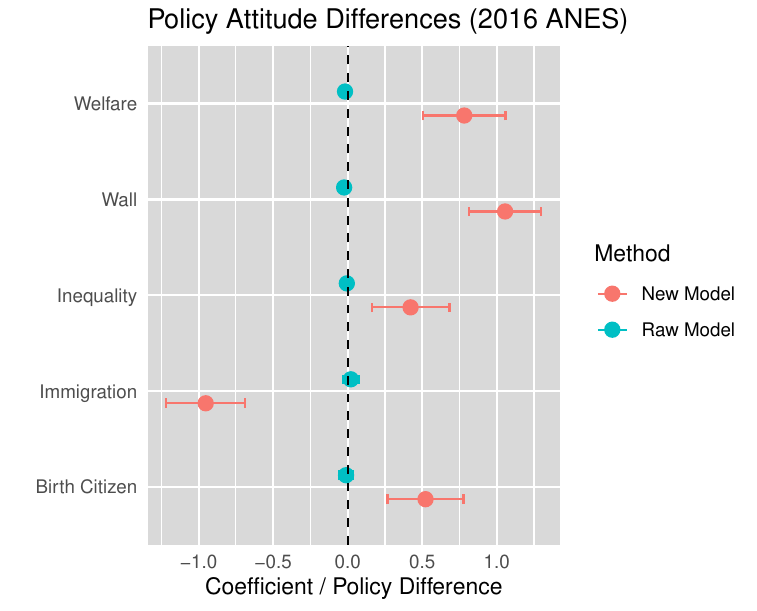}
        \caption*{(a) 2016 ANES}
    \end{minipage}
    \hfill
    \begin{minipage}{0.42\textwidth}
        \centering
        \includegraphics[width=\textwidth]{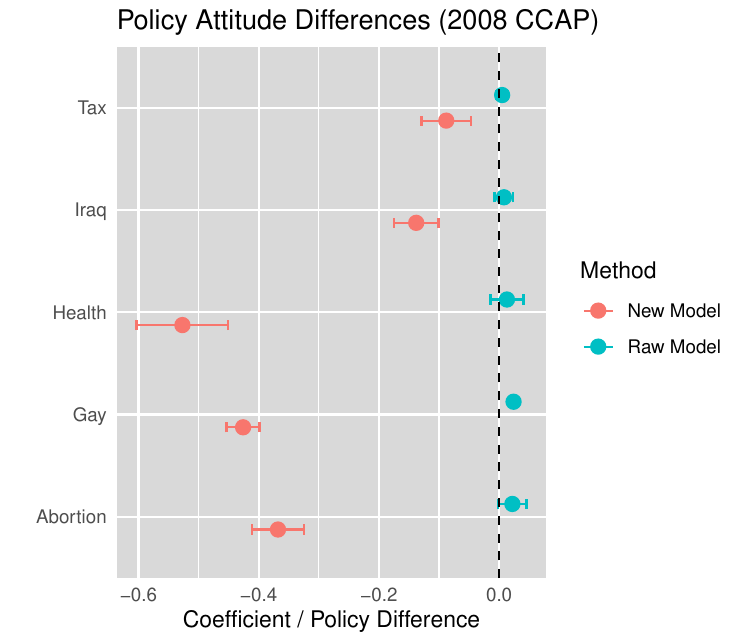}
        \caption*{(b) 2008 CCAP}
    \end{minipage}

    \vspace{1em}

    % Bottom row
    \begin{minipage}{0.48\textwidth}
        \centering
        \includegraphics[width=\textwidth]{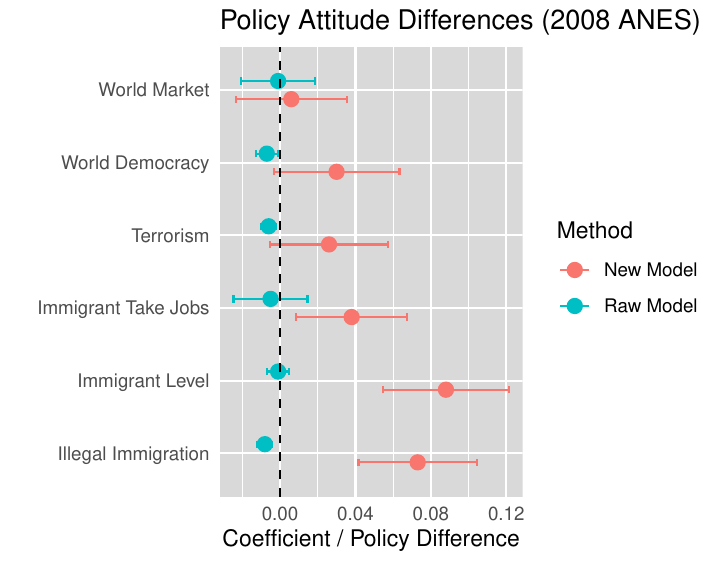}
        \caption*{(c) 2008 ANES}
    \end{minipage}
    \hfill
    \begin{minipage}{0.48\textwidth}
        \centering
        \includegraphics[width=\textwidth]{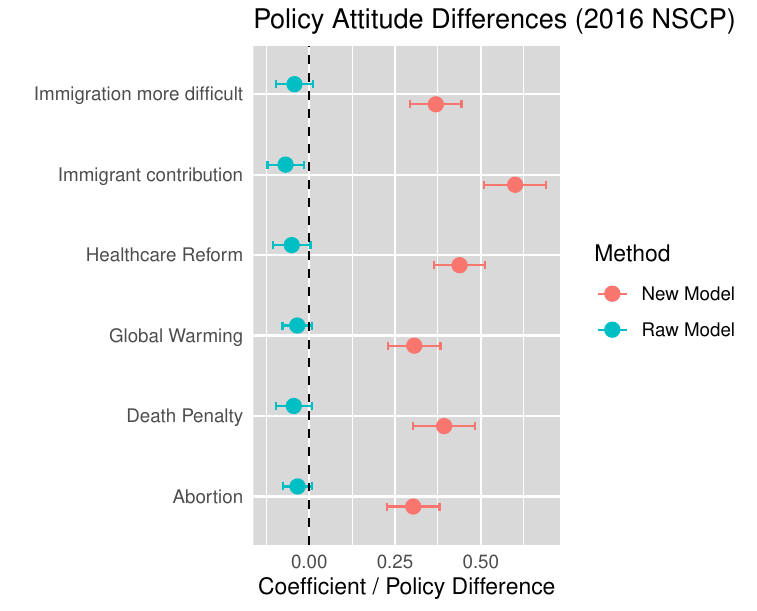}
        \caption*{(d) 2016 NSCP}
    \end{minipage}

    \label{fig:policy_diff}
\end{figure}

Figure~\ref{fig:policy_diff} presents estimated Black–White differences in policy attitudes implied by authoritarianism using two measurement strategies: conventional multi-item scales and bias-corrected latent scores from partial scalar MG-CFA models. Across policy domains—immigration, labor market competition, terrorism, and global attitudes—the conventional scale yields estimates close to zero, with confidence intervals largely overlapping zero.

In contrast, latent authoritarianism produces larger and more consistent Black–White differences across datasets and outcomes. This pattern indicates that measurement non-invariance in the child-rearing items attenuates conventional estimates of authoritarianism’s political implications, rather than reflecting substantively weak associations. 

Some variation across policy outcomes reflects differences in survey design and item context rather than instability in the underlying latent structure–outcome relationship. The 2008 CCAP exhibits relatively larger threshold differences for certain items, consistent with heterogeneity in item interpretation and response behavior across survey contexts \citep{Feldman1997, Pietryka2022}. However, even in this case, partial scalar MG-CFA recovers a stable latent structure and yields substantively comparable patterns in policy relationships.

Taken together, results from the 2008 and 2016 ANES, the 2008 CCAP, and the 2016 NSCP indicate that conventional multi-item child-rearing scales attenuate the relationship between authoritarianism and policy preferences. Accounting for threshold non-invariance produces stronger and more consistent associations across domains and datasets.

\section{Discussion \& Conclusion}

This study revisits a foundational question in the study of racial politics: whether observed group differences in authoritarianism reflect genuine variation in underlying predispositions or artifacts of measurement. Using a multi-group ordinal framework with anchor-based identification, this study shows that widely used racial comparisons of authoritarianism are sensitive to how item thresholds and response processes are modeled. The central substantive implication is that estimated racial differences in authoritarianism are not invariant to measurement structure. When threshold heterogeneity is accounted for, the magnitude—and in some specifications the interpretation—of racial gaps differs meaningfully from conventional estimates based on summed scales or standard multi-group factor models.

This result has direct implications for the study of political psychology. The child-rearing authoritarianism battery has long been used to infer stable group differences in political dispositions and to explain downstream variation in ideology and policy preferences. However, if item thresholds differ across racial groups, respondents with equivalent latent predispositions may systematically select different response categories. Under these conditions, observed differences in scale distributions do not map cleanly onto differences in authoritarian predispositions. Instead, part of what has been interpreted as group divergence reflects variation in how respondents use ordinal response scales.

A second implication concerns the interpretation of ideological structure across groups. Because measurement precision and threshold structure vary across racial groups, conventional models may differentially attenuate relationships between authoritarianism and political attitudes. This can create the appearance of weaker ideological constraint in one group relative to another, even when the underlying latent relationships are more similar than observed estimates suggest. What appears to be substantive heterogeneity in political psychology may therefore reflect systematic differences in measurement properties rather than differences in political structure.

More broadly, these findings speak to a central challenge in the study of racial politics: separating substantive attitudinal divergence from measurement artifacts embedded in survey instruments. A large body of work on racial differences in authoritarianism, racial resentment, efficacy, and related constructs relies on short ordinal batteries that impose strong and often untested comparability assumptions across groups. The results presented here demonstrate that once these assumptions are relaxed in a principled manner, cross-group comparisons become explicitly model-dependent. In particular, latent mean differences are identified only under clearly specified and substantively defensible anchor constraints, and alternative plausible measurement specifications can meaningfully alter substantive conclusions.

Importantly, these results do not imply that prior findings are invalid or that group differences are illusory. Rather, they underscore that both the magnitude and interpretation of these differences depend on measurement assumptions that are often left implicit. By incorporating anchor items and explicitly modeling threshold non-invariance, researchers can more clearly separate latent construct differences from response-scale artifacts. This improves substantive inference by making identifying assumptions transparent rather than implicit.

These implications extend beyond authoritarianism. Cross-group comparisons are central to research on ideology, affective polarization, trust, resentment, and democratic norms. In each domain, scholars routinely compare scale means and regression coefficients across racial and ethnic groups. If measurement structures differ systematically across groups, then observed differences in levels and associations may partly reflect properties of survey instruments rather than properties of political attitudes. The present results suggest that careful attention to item-level invariance is a prerequisite for credible cross-group inference in political behavior research.

Future work should extend this framework to other widely used political attitude batteries and explore the sources of threshold noninvariance. Differences in linguistic interpretation, normative evaluation, and culturally structured response styles may all contribute to heterogeneous item functioning. Understanding these mechanisms is not only a measurement concern but also a substantive question about political socialization and identity formation.

Ultimately, debates about racial differences in authoritarianism are also debates about how political attitudes are measured and compared. By clarifying the identification conditions required for valid cross-group inference, this study contributes to a more transparent foundation for research on racial politics and political psychology.

\clearpage

\begin{singlespace}

\end{singlespace}

\newpage

\renewcommand{\thefigure}{A\arabic{figure}}
\setcounter{figure}{0}

\begin{center}
\section*{Appendix A: \\ Anchor Items}
\end{center}

This appendix provides additional details on anchor item selection, measurement invariance tests, bootstrap procedures, and the Monte Carlo simulation used to evaluate estimator performance.

Table A1 summarizes the datasets used in the analysis and the corresponding anchor item sets employed for partial scalar identification.

\begin{singlespace}
\renewcommand{\arraystretch}{1.1}
\begin{table}[H]
\centering
\caption{Summary of Anchor Item Sources Across Datasets}
\label{tab:anchor_summary}
\begin{tabular}{llcp{6cm}}
\toprule
Dataset & Anchor Type & Items & Rationale \\
\midrule
2008 ANES & Cognitive engagement items & 4 & Conceptually independent of authoritarianism. \\
2008 CCAP & Optimism / pessimism & 4 & Psychologically neutral affective orientation. \\
2016 NSCP & Feeling thermometers & 4 & Social evaluations unrelated to child-rearing values. \\
2016 ANES & Personality descriptors & 4 & Trait-based self-descriptions. \\
\bottomrule
\end{tabular}
\end{table}
\end{singlespace}

The correlation matrices reported below confirm that the anchor items are only weakly related to one another. This pattern is expected and desirable: anchor items are not intended to measure a common latent construct but instead function as independent reference points that stabilize identification in partial scalar models.

This pattern is expected and desirable. Anchor items are not intended to measure a common latent construct; rather, they function as independent reference indicators that stabilize identification in partial scalar invariance models. Weak correlations therefore suggest that the anchors do not introduce an unintended latent dimension into the measurement model and instead serve their intended role as neutral reference points.

The following tables report correlation matrices for the anchor items used in each dataset.

\subsection*{\centering Anchor Items: 2008 ANES}

\begin{enumerate}
    \item Does the respondent prefer simple problems or complex problems?
    \item Does the respondent like or dislike responsibility for thinking?
    \item Does the respondent enjoy taking responsibility for handling complex thinking?
    \item To what extent can people change the kind of person they are?

\end{enumerate}

\setcounter{table}{0} % Reset table counter for the appendix
\renewcommand{\thetable}{A\arabic{table}} % Define new numbering format (A1, A2, etc.)

\begin{table}[h!]
    \centering
    \caption{Correlation Matrix between Anchor Items in the 2008 ANES data}
    \label{tab:correlation_matrix}
    \begin{tabular}{lccccc}
        \toprule
        \ & Anchor 1 & Anchor 2 & Anchor 3 & Anchor 4\\
        \midrule
        Anchor 1 & 1.000 &       &       &       &       \\
        Anchor 2 & 0.257 & 1.000 &       &       &       \\
        Anchor 3 & -0.392 & -0.285 & 1.000 &       &       \\
        Anchor 4 & -0.126 & -0.166 & 0.112 & 1.000 &       \\

        \bottomrule
    \end{tabular}
\end{table}

\begin{center}
    \subsection*{Anchor Items: 2008 CCAP}
\end{center}

We use optimism–pessimism items from the CCAP personality and psychological profile battery. These items are drawn from the Life Orientation Test–Revised (LOT-R), a widely used measure of dispositional optimism in psychology.

Anchor Item Questions:

\begin{enumerate}
    \item It's easy for me to relax [MCAP917].
    \item I enjoy my friends a lot [MCAP920]. 
    \item It's important for me to keep busy [MCAP921]. 
    \item I don't get upset too easily [MCAP923] 

\end{enumerate}

\begin{singlespace}
\begin{table}[H]
    \centering
    \caption{Correlation Matrix between Anchor Items in the 2008 CCAP data}
    \label{tab:correlation_matrix}
    \begin{tabular}{lrrrrr}
        \toprule
        & Anchor 1 & Anchor 2 & Anchor 3 & Anchor 4 \\
        \midrule
Anchor 1 & 1.00 & & & \\ 
  Anchor 2 & 0.16 & 1.00 & & & \\ 
  Anchor 3 & -0.03 & 0.15 & 1.00 & \\ 
  Anchor 4 & 0.33 & 0.14 & 0.05 & 1.00 \\ 
        \bottomrule
    \end{tabular}
\end{table}
\end{singlespace}

Although some psychological constructs exhibit substantial measurement non-invariance across racial groups, this pattern is not universal. For example, optimism and pessimism scales, such as the LOT-R–style items included in the CCAP, consistently demonstrate full measurement invariance across Black and White respondents. These items capture a basic affective orientation--general expectations about the likelihood of good or bad outcomes--that is not culturally or normatively loaded. As a result, respondents across racial groups interpret these statements in similar ways, producing stable loadings and thresholds. The contrast between these invariant optimism items and the authoritarianism items analyzed in this study highlights that measurement bias is construct-specific rather than group-specific: racial non-invariance is not an inevitable artifact of survey response patterns among Black Americans but arises primarily for value-laden constructs shaped by racialized socialization and historical experience. This comparison reinforces the substantive conclusion of this article—namely, that observed racial differences in authoritarianism reflect genuine latent variation rather than systematic misunderstanding of item wording.

\begin{center}
\subsection*{Anchor Items: 2016 NSCP Data}
\end{center}

For the 2016 NSCP data, we chose the following four  items as anchor items:

\begin{enumerate}
 \item Feeling thermometer for Latino people
 \item Feeling thermometer for Asian people
 \item Feeling thermometer for Jewish people
 \item Feeling thermometer for the alt-right movement

\end{enumerate}

The feeling thermometer scores range from 0 to 100. For analysis, we recoded them into four equal intervals (0--24, 25--49, 50--74, 75--100). This recoding facilitates stable estimation of thresholds for partial scalar identification while preserving substantive meaning: each category represents a distinct level of affect toward the target group. Coarse categorization also mitigates sparsity in the tails of the distribution and ensures that anchor items function reliably across racial groups, without affecting their role as independent reference points \citep{Steenkamp1998, Vandenberg2000, Davidov2014}.

\begin{singlespace}
\begin{table}[H]
    \centering
    \caption{Correlation Matrix between Anchor Items in the 2016 NSCP data}
    \label{tab:correlation_matrix}
    \begin{tabular}{lrrrr}
        \toprule
        & Anchor 1 & Anchor 2 & Anchor 3 & Anchor 4 \\
        \midrule
Anchor 1 & 1.000  \\ 
Anchor 2 & -0.104 & 1.000 \\ 
Anchor 3 & -0.101 & 0.527 & 1.000 \\ 
Anchor 4 & -0.006 & 0.005 & 0.034 & 1.000 \\ 
        \bottomrule
    \end{tabular}
\end{table}
\end{singlespace}

Anchor items should ideally be conceptually unrelated to the focal construct. In the CCAP, we used affective personality items (optimism/pessimism) that are neutral and invariant. In the 2016 NSCP, no comparable neutral items were available; we therefore used feeling thermometer items toward various social groups. While socially salient, these items are conceptually distinct from the child-rearing authoritarianism items and are not indicators of authoritarianism. Invariance tests confirm that they function equivalently across racial groups, providing a valid basis for partial scalar identification. 

\begin{center}
\subsection*{Anchor Items: 2016 ANES}
\end{center}

\begin{enumerate}
    \item The respondent has many more opinions than the average person [v162252].
    \item Please mark how well the following pair of words describes you, even if one word describes you better than the other: “sympathetic, warm.” Forward/reverse response option order [v162339].
    \item Please mark how well the following pair of words describes you, even if one word describes you better than the other: “conventional, uncreative.” Forward/reverse response option order [v162342].
    \item The respondent likes to have strong opinions even when not personally involved [v162248].
\end{enumerate}

Because the original items used different response formats (five- and seven-point scales), all variables were recoded into a common three-category ordinal scale to ensure comparability across indicators. Substantive results are unchanged when the original response scales are used.

\begin{singlespace}
\begin{table}[H]
    \centering
    \caption{Correlation Matrix between Anchor Items in the 2016 ANES data}
    \label{tab:correlation_matrix}
    \begin{tabular}{lrrrr}
        \toprule
        & Anchor 1 & Anchor 2 & Anchor 3 & Anchor 4 \\
        \midrule
Anchor 1 & 1.000  \\ 
Anchor 2 & -0.0494 & 1.000 \\ 
Anchor 3 & -0.0608 & -0.052 & 1.000 \\ 
Anchor 4 & -0.0905 & 0.0760 & 0.1217 & 1.000 \\ 
        \bottomrule
    \end{tabular}
\end{table}
\end{singlespace}

Note that low correlations among anchor items are intentional, as these items serve as independent reference points for partial scalar identification rather than indicators of a common latent factor.

\newpage

\begin{center}
\section*{Appendix B: \\MG-CFA Partial Measurement Invariance Results}
\end{center}

\begin{singlespace}

\begin{table}[H]
\centering
\footnotesize
\caption{MG-CFA Results for Authoritarianism and Anchor Items (2008 ANES)}
\label{tab:cfa_appendix}
\begin{tabular}{lcccc|cccc}
\hline
& \multicolumn{4}{c|}{White} & \multicolumn{4}{c}{Black} \\
                       & Std. Est. & p-value & Threshold & Residual & Std. Est. & p-value & Threshold & Residual \\
\hline
auth\_1   & 0.533 & 0.000 & -0.864 & 0.716 & 0.395 & 0.000 & -1.127 & 0.496 \\
auth\_2   & 0.664 & 0.000 & -0.556 & 0.559 & 0.617 & 0.000 & -0.912 & 0.231 \\
auth\_3   & 0.594 & 0.000 & -0.223 & 0.647 & 0.729 & 0.000 & -0.481 & 0.101 \\
auth\_4   & 0.566 & 0.000 & 0.103  & 0.679 & 0.669 & 0.000 & 0.214  & 0.128 \\
anchor1   & 0.607 & 0.000 & -0.804 & 0.632 & 0.307 & 0.000 & -0.717 & 1.137 \\
anchor2   & -0.527 & 0.000 & -1.160 & 0.722 & -0.272 & 0.000 & -1.055 & 1.121 \\
anchor3   & 0.185 & 0.000 & -0.976 & 0.966 & 0.085 & 0.000 & -0.789 & 1.518 \\
anchor4   & -0.457 & 0.000 & 0.036  & 0.791 & -0.378 & 0.302 & 0.053  & 0.405 \\
\hline \hline
\end{tabular}
\end{table}

\begin{table}[H]
\centering
\footnotesize
\caption{MG-CFA Results for Authoritarianism and Anchor Items (2008 CCAP)}
\label{tab:cfa_appendix_new}
\begin{tabular}{lcccc|cccc}
\hline
& \multicolumn{4}{c|}{White} & \multicolumn{4}{c}{Black} \\
                       & Std. Est. & p-value & Threshold & Residual & Std. Est. & p-value & Threshold & Residual \\
\hline
auth\_1      & 0.767 & 0.000 & 0.711 & 0.411 & 0.675 & 0.000 & 0.508 & 1.068 \\
auth\_2      & 0.759 & 0.000 & 0.206 & 0.423 & 0.685 & 0.000 & 0.151 & 0.990 \\
auth\_3      & -0.758 & 0.000 & 0.253 & 0.426 & -0.736 & 0.000 & 0.200 & 0.738 \\
auth\_4      & 0.523 & 0.000 & -0.522 & 0.727 & 0.560 & 0.000 & -0.454 & 0.906 \\
anchor1     & -0.035 & 0.006 & 0.259 & 0.999 & -0.058 & 0.006 & 0.351 & 0.541 \\
anchor2     & 0.039 & 0.007 & 0.550 & 0.998 & 0.041 & 0.007 & 0.472 & 1.354 \\
anchor3     & -0.044 & 0.007 & 1.025 & 0.998 & -0.051 & 0.007 & 0.950 & 1.162 \\
anchor4     & -0.056 & 0.000 & 0.148 & 0.997 & -0.092 & 0.000 & 0.195 & 0.567 \\
\hline \hline
\end{tabular}
\end{table}

\begin{table}[H]
\centering
\footnotesize
\caption{MG-CFA Results for Authoritarianism and Anchor Items (2016 ANES)}
\label{tab:cfa_appendix_new2}
\begin{tabular}{lcccc|cccc}
\hline
& \multicolumn{4}{c|}{White} & \multicolumn{4}{c}{Black} \\
                       & Std. Est. & p-value & Threshold & Residual & Std. Est. & p-value & Threshold & Residual \\
\hline
auth\_1      & 0.836 & 0.000 & -0.516 & 0.300 & 0.637 & 0.000 & -0.516 & 1.129 \\
auth\_2      & 0.857 & 0.000 & -0.218 & 0.265 & 0.806 & 0.000 & -0.218 & 0.439 \\
auth\_3      & 0.739 & 0.000 & 0.160 & 0.454 & 0.807 & 0.000 & 0.160 & 0.322 \\
auth\_4      & 0.541 & 0.000 & 0.588 & 0.707 & 0.573 & 0.000 & 0.593 & 0.659 \\

anchor1      & -0.167 & 0.001 & -0.036 & 0.972 & -0.083 & 0.001 & -0.017 & 4.449 \\
anchor2      & 0.053 & 0.344 & -1.482 & 0.997 & 0.057 & 0.344 & -1.508 & 0.963 \\
anchor3      & 0.290 & 0.000 & 0.289 & 0.916 & 0.360 & 0.000 & 0.341 & 0.626 \\
anchor4      & -0.345 & 0.000 & -1.919 & 0.881 & -0.426 & 0.000 & -2.256 & 0.592 \\
\hline \hline
\end{tabular}
\end{table}

\begin{table}[H]
\centering
\footnotesize
\caption{MG-CFA Results for Authoritarianism and Anchor Items (2016 NSCP)}
\label{tab:cfa_groups_new}
\begin{tabular}{lcccc|cccc}
\hline
& \multicolumn{4}{c|}{White} & \multicolumn{4}{c}{Black} \\
                       & Std. Est. & p-value & Threshold & Residual & Std. Est. & p-value & Threshold & Residual \\
\hline
auth\_1      & 0.558 & 0.000 & -0.452 & 0.688 & 0.541 & 0.000 & -0.452 & 0.713 \\
auth\_2      & 0.607 & 0.000 & -0.056 & 0.631 & 0.521 & 0.000 & -0.049 & 0.939 \\
auth\_3      & 0.536 & 0.000 & 0.473 & 0.713 & 0.342 & 0.000 & 0.310 & 2.054 \\
auth\_4      & 0.509 & 0.000 & 0.469 & 0.741 & 0.407 & 0.000 & 0.385 & 1.238 \\
anchor1      & -0.810 & 0.000 & -1.649 & 0.345 & -0.877 & 0.000 & -1.833 & 0.187 \\
anchor2      & -0.818 & 0.000 & -1.929 & 0.330 & -0.820 & 0.000 & -1.985 & 0.309 \\
anchor3      & -0.671 & 0.000 & -2.022 & 0.550 & -0.693 & 0.000 & -2.144 & 0.462 \\
anchor4      & 0.256 & 0.000 & 0.137 & 0.934 & 0.272 & 0.000 & 0.150 & 0.778 \\
\hline \hline
\end{tabular}
\end{table}

\end{singlespace}

We estimated a MG-CFA to compare latent authoritarianism across racial groups. Full measurement invariance, that is, equality of loadings and thresholds for all items, did not hold, as several anchor items showed differential functioning across groups. To address this, we implemented partial measurement invariance, allowing anchor items to vary while constraining the remaining items. This approach preserves the comparability of the latent factor across groups, ensures valid estimation of latent mean differences, and aligns with standard psychometric practice when full invariance is unrealistic \citep{Byrne1989, Millsap2004}. 

It is important to note that including anchor items in a one-factor model often reduces conventional model fit indices. This occurs because anchor items are conceptually distinct from the authoritarianism items and are not intended to load on the same latent dimension. Their primary role is not to improve model fit, but to facilitate the identification and accurate estimation of latent mean differences across groups.

\newpage

\begin{center}
\section*{Appendix C: \\ Monte Carlo Simulation Procedure}
\end{center}

The Monte Carlo simulation is designed to directly compare multi-item scales with latent variable models under both anchor-based partial scalar invariance and full scalar invariance specifications. By varying the magnitude of measurement non-invariance (DIF), we assess the extent to which non-invariance biases estimates of group differences and whether partial scalar invariance models can recover latent mean differences without bias.

In addition, we vary factor loadings, residual variances, and the number of response categories to capture a range of realistic measurement conditions and signal-to-noise ratios. This design allows us to evaluate the robustness of alternative estimators under varying levels of measurement quality. The simulation proceeds as follows:

\paragraph{1. Simulation design.} 
We generated $R = 500$ Monte Carlo replications, each with $N = 1{,}000$ respondents equally split between two groups: White and Black respondents. The latent trait of interest, authoritarianism, was assumed to follow a normal distribution with a true group mean difference of $\Delta = 0.2$ favoring Black respondents. Specifically, for individual $i$ in group $g$ ($g = 0$ for White, $g = 1$ for Black):
\[
\eta_i \sim \mathcal{N}(\mu_\eta^{(g)}, 1), \quad \mu_\eta^{(0)} = 0, \quad \mu_\eta^{(1)} = \Delta.
\]

\paragraph{2. Item generation.} 
Eight items were simulated, consisting of four DIF items and four anchor items. The continuous item response for item $j$ of individual $i$ was generated as:
\[
y_{ij} = \lambda_j \eta_i + \epsilon_{ij}, \quad \epsilon_{ij} \sim \mathcal{N}(0, \sigma^2_\epsilon).
\]

DIF was introduced for the first four items via a group-specific intercept shift:
\[
y_{ij} \leftarrow y_{ij} + \delta_j \cdot g,
\]
where $\delta_j$ varies across simulations (0.05 to 0.9), and the remaining four anchor items are invariant across groups ($\delta_j = 0$).

\paragraph{3. Discretization of responses.} 
To reflect the ordinal nature of survey responses, continuous item responses were discretized into four ordered categories using fixed thresholds:
\[
\text{category}_{ij} = 
\begin{cases}
1 & y_{ij} \leq -1 \\
2 & -1 < y_{ij} \leq 0 \\
3 & 0 < y_{ij} \leq 1 \\
4 & y_{ij} > 1
\end{cases}.
\]

\paragraph{4. Estimation procedures.} 
For each replication, three estimators of group differences are computed:

\begin{enumerate}
\item \textbf{Multi-item scale}: Summed scores across the four DIF items, averaged within each group. The group difference is computed as the mean difference between Black and White respondents.

\item \textbf{Partial scalar invariance CFA (anchor-based)}: A one-factor CFA is estimated using all eight items, constraining loadings and thresholds for anchor items while freeing thresholds for DIF items. Latent means are estimated for each group, and the Black–White difference is computed.

\item \textbf{Full scalar invariance CFA}: A one-factor CFA is estimated using all eight items, imposing equality constraints on factor loadings and thresholds across groups. No parameters are freed across groups. Latent means are then estimated for each group, and the Black–White difference is computed under full scalar invariance.
\end{enumerate}

\subsection*{Partial \& Full Scalar Invariance vs. Multi-Item}

\begin{figure}[H]
 \caption{Evaluating Measurement Bias $\lambda=0.8$}
    \centering
          \includegraphics[width=1.0\textwidth]{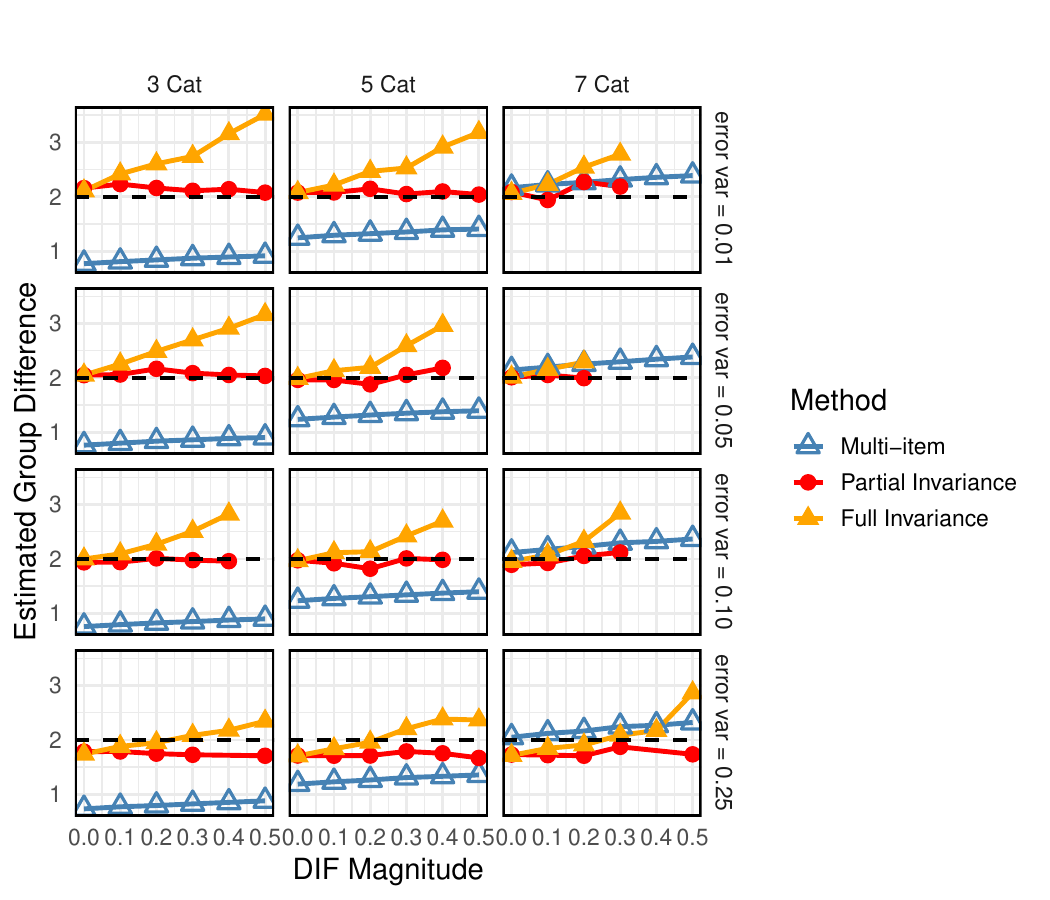}
    \label{app:sim-1}
\end{figure}

\begin{figure}[H]
    \centering
        \caption{Evaluating Measurement Bias $\lambda=0.9$}

          \includegraphics[width=1.0\textwidth]{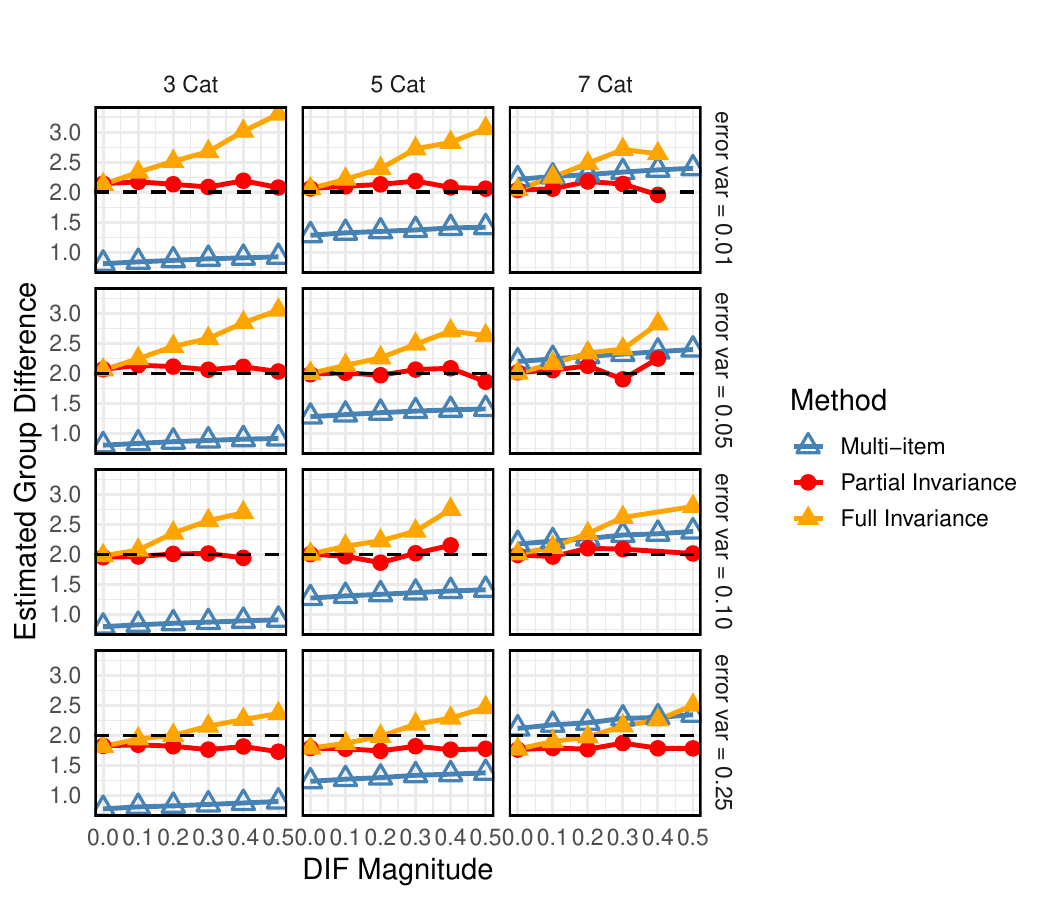}

    \label{app:sim-2}
\end{figure}

\begin{figure}[H]
    \centering
        \caption{Evaluating Measurement Bias $\lambda=1.0$}

          \includegraphics[width=1.0\textwidth]{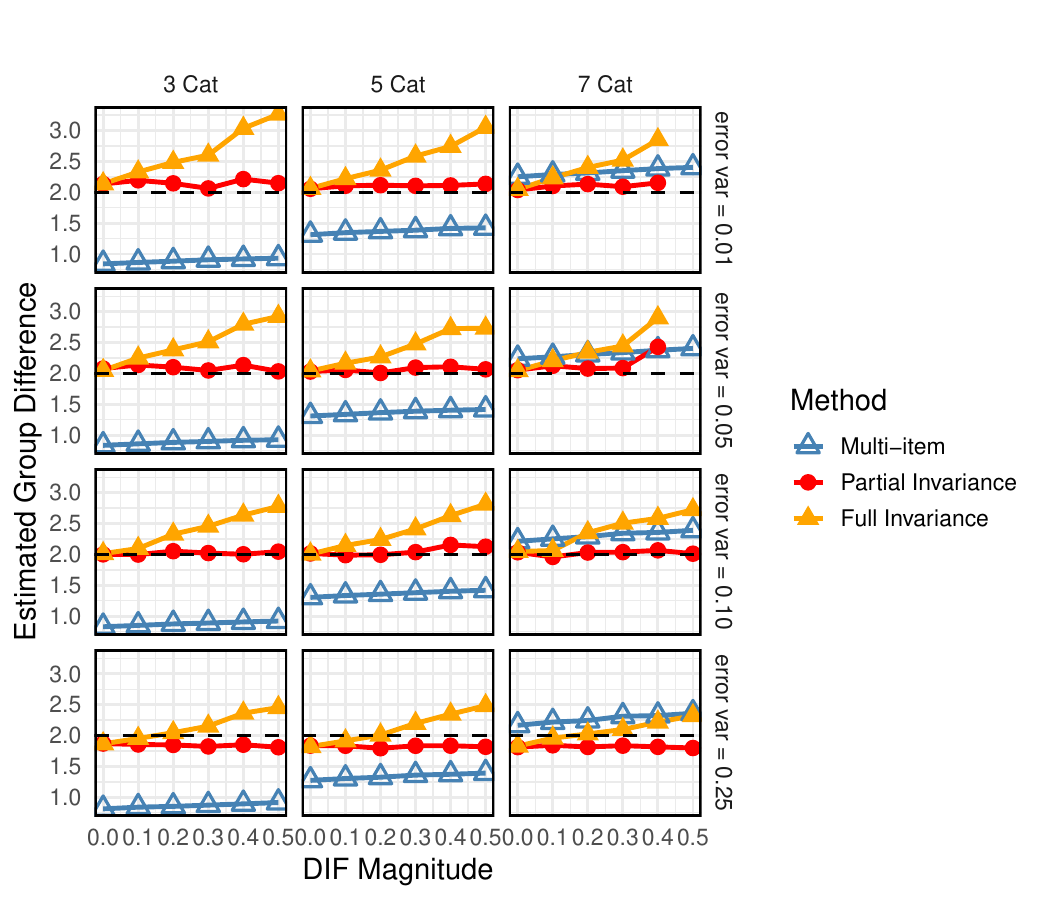}

    \label{app:sim-3}
\end{figure}

Figures~\ref{app:sim-1}, \ref{app:sim-2}, and \ref{app:sim-3} show that anchor-based partial invariance, on average, outperforms both the full invariance specification and the multi-item scale method. Across all category structures, the anchor-based model more accurately recovers the true group difference of 2.0 when factor loadings are high and error variances are low.

In contrast, when error variance is high, the performance of all methods converges. In this setting, the full invariance model exhibits a slight advantage in some conditions. This likely reflects a bias–variance trade-off: when measurement error is large, the benefits of modeling item-specific non-invariance diminish, while the stronger cross-group equality constraints imposed by the full invariance specification can yield slightly more stable estimates.

Furthermore, as the number of response categories increases, the bias across methods decreases, indicating that improved measurement granularity enhances the precision of group difference estimation. In addition, when factor loadings are weak, the partial invariance model often encounters convergence issues and fails to produce stable estimates. In contrast, when factor loadings exceed 0.8, the partial invariance approach consistently outperforms the multi-item scale, yielding estimates close to the true value. Moreover, a greater number of response categories facilitates the detection of group differences and leads to more stable and consistent estimates.

\subsection*{\centering Evaluating Measurement Bias with 8 Items}

We also conduct a simulation design that deliberately distinguishes between the number of items used for scale construction and the number of items used for measurement modeling in order to mimic common empirical practice and isolate sources of bias. Specifically, while the full data-generating process contains eight items (four invariant anchors and four DIF items), we evaluate multi-item aggregation and full scalar invariance using only the four non-anchor (DIF) items. This setup reflects a common applied strategy in which researchers rely on a short subset of widely used child-rearing items without explicitly modeling measurement non-invariance. In contrast, the anchor-based partial scalar invariance model uses the full eight-item structure, with four items serving as invariant anchors to identify the latent scale and four items allowed to exhibit group-specific thresholds. This design allows us to directly compare a naïve aggregation strategy, a misspecified full invariance model, and a correctly specified partial invariance model under identical underlying latent conditions. By holding the data-generating process constant while varying the estimation strategy, we are able to isolate how much bias is attributable to ignoring DIF versus imposing overly restrictive invariance assumptions.

We further show that the main results are robust to using the full set of eight items. As expected, incorporating more items improves estimation stability and reduces sampling variability and bias, particularly for the latent variable models. Importantly, however, the substantive pattern of results remains unchanged: partial scalar invariance consistently recovers the true group differences most accurately, while full scalar invariance and multi-item aggregation continue to exhibit bias under increasing levels of non-invariance. This reinforces the central conclusion that the key driver of bias is not simply the number of items, but whether measurement non-invariance is explicitly modeled. \\

\begin{figure}[H]
    \centering
        \caption{Evaluating Measurement Bias (8 Items)}

          \includegraphics[width=1.0\textwidth]{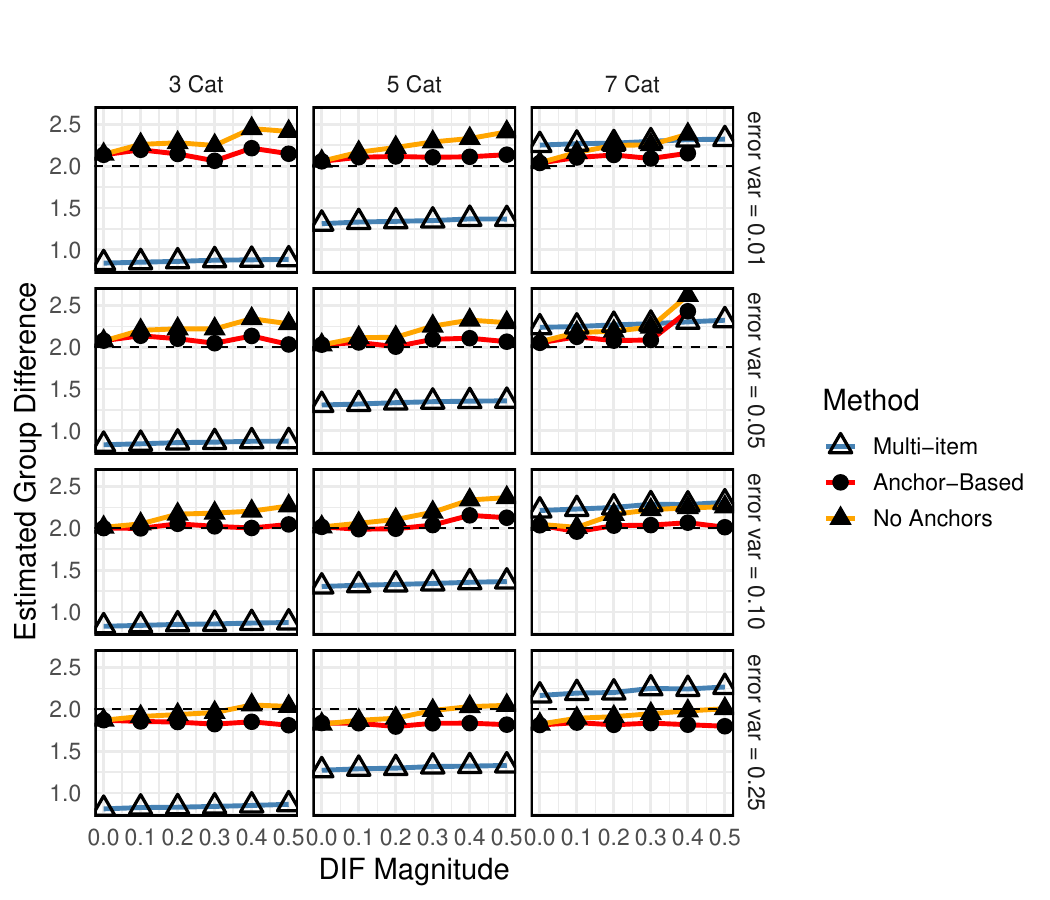}

    \label{app:sim-8item}
\end{figure}

\newpage

\section*{Appendix E: Construction of Threshold Difference Estimates}

Let $\tau_{jk}^{(g)}$ denote the estimated threshold for item $j$ at category $k$ in group $g \in \{\text{White}, \text{Black}\}$. Threshold parameters are obtained from the fitted multi-group CFA model with ordered categorical indicators.

Threshold differences are computed as:

\begin{equation}
\Delta \tau_{jk} = \tau_{jk}^{(Black)} - \tau_{jk}^{(White)}.
\end{equation}

Positive values indicate that Black respondents require higher levels of the latent trait to endorse category $k$, whereas negative values indicate lower thresholds relative to White respondents.

All threshold parameters are extracted from the model-implied parameter estimates under the partial scalar invariance specification described in the main text. No post-estimation rescaling is applied.

\begin{table}[H]
\centering
\scriptsize
\caption{Estimated Thresholds (2016 NSCP)}
\label{tab:thresholds_full}

\begin{tabular}{llcccc}
\hline
Item & Threshold & Group & Estimate & SE & 95\% CI \\
\hline

\multicolumn{6}{c}{\textbf{Child-Rearing Items}} \\
\hline

auth\_1 & t1 & White & -0.432 & 0.020 & [-0.471, -0.392] \\
auth\_2 & t1 & White & -0.026 & 0.020 & [-0.065, 0.012] \\
auth\_3 & t1 & White & 0.496  & 0.020 & [0.455, 0.536] \\
auth\_4 & t1 & White & 0.502  & 0.020 & [0.462, 0.542] \\

auth\_1 & t1 & Black & -0.730 & 0.070 & [-0.868, -0.593] \\
auth\_2 & t1 & Black & -0.474 & 0.066 & [-0.604, -0.344] \\
auth\_3 & t1 & Black & -0.055 & 0.064 & [-0.180, 0.070] \\
auth\_4 & t1 & Black & -0.036 & 0.064 & [-0.161, 0.089] \\

\hline
\multicolumn{6}{c}{\textbf{Anchor Items}} \\
\hline

anchor1 & t1 & White & -1.650 & 0.033 & [-1.714, -1.585] \\
anchor1 & t2 & White & -0.714 & 0.021 & [-0.756, -0.672] \\
anchor1 & t3 & White & 0.034  & 0.020 & [-0.005, 0.072] \\

anchor1 & t1 & Black & -1.765 & 0.117 & [-1.994, -1.536] \\
anchor1 & t2 & Black & -1.015 & 0.077 & [-1.166, -0.864] \\
anchor1 & t3 & Black & -0.239 & 0.064 & [-0.365, -0.112] \\

anchor2 & t1 & White & -1.936 & 0.041 & [-2.016, -1.856] \\
anchor2 & t2 & White & -0.939 & 0.023 & [-0.984, -0.893] \\
anchor2 & t3 & White & -0.142 & 0.020 & [-0.180, -0.103] \\

anchor2 & t1 & Black & -1.866 & 0.126 & [-2.113, -1.619] \\
anchor2 & t2 & Black & -1.026 & 0.078 & [-1.178, -0.874] \\
anchor2 & t3 & Black & -0.225 & 0.064 & [-0.351, -0.099] \\

anchor3 & t1 & White & -2.046 & 0.045 & [-2.134, -1.958] \\
anchor3 & t2 & White & -1.032 & 0.024 & [-1.079, -0.986] \\
anchor3 & t3 & White & -0.320 & 0.020 & [-0.359, -0.281] \\

anchor3 & t1 & Black & -1.904 & 0.130 & [-2.159, -1.650] \\
anchor3 & t2 & Black & -1.015 & 0.077 & [-1.166, -0.864] \\
anchor3 & t3 & Black & -0.299 & 0.065 & [-0.426, -0.172] \\

anchor4 & t1 & White & 0.117  & 0.020 & [0.078, 0.155] \\
anchor4 & t2 & White & 0.791  & 0.022 & [0.748, 0.834] \\
anchor4 & t3 & White & 1.406  & 0.028 & [1.350, 1.462] \\

anchor4 & t1 & Black & 0.340  & 0.065 & [0.212, 0.468] \\
anchor4 & t2 & Black & 0.912  & 0.074 & [0.766, 1.058] \\
anchor4 & t3 & Black & 1.538  & 0.100 & [1.341, 1.735] \\

\hline
\end{tabular}
\end{table}

\begin{table}[H]
\centering
\caption{Threshold Differences (2016 NSCP)}
\label{tab:nscp16_thresholds}
\small

\begin{tabular}{lcccc}
\hline
Item & t1 & t2 & t3 & Type \\
\hline

Item 1   & -0.432 / -0.730 & -- & -- & Child-rearing \\
Item 2   & -0.026 / -0.474 & -- & -- & Child-rearing \\
Item 3  &  0.496 / -0.055 & -- & -- & Child-rearing \\
Item 4  &  0.502 / -0.036 & -- & -- & Child-rearing \\
\hline

Anchor 1 & -1.650 / -1.765 & -0.714 / -1.015 &  0.034 / -0.239 & Anchor \\
Anchor 2 & -1.936 / -1.866 & -0.939 / -1.026 & -0.142 / -0.225 & Anchor \\
Anchor 3 & -2.046 / -1.904 & -1.032 / -1.015 & -0.320 / -0.299 & Anchor \\
Anchor 4 &  0.117 /  0.340 &  0.791 /  0.912 &  1.406 /  1.538 & Anchor \\
\hline

\end{tabular}

{\footnotesize
\par\noindent\raggedright
\textit{Note:} Values are reported as White / Black respondents.
}

\end{table}

\subsection*{2008 CCAP}

\begin{table}[H]
\centering
\scriptsize
\caption{Estimated Thresholds (2008 CCAP)}
\label{tab:thresholds_full_anes16}

\begin{tabular}{llcccc}
\hline
Item & Threshold & Group & Estimate & SE & 95\% CI \\
\hline

\multicolumn{6}{c}{\textbf{Child-Rearing Items}} \\
\hline

auth\_1 & t1 & White & -0.524 & 0.045 & [-0.613, -0.435] \\
auth\_2 & t1 & White & -0.215 & 0.043 & [-0.300, -0.130] \\
auth\_3 & t1 & White &  0.164 & 0.043 & [0.079, 0.249] \\
auth\_4 & t1 & White &  0.607 & 0.046 & [0.517, 0.698] \\

auth\_1 & t1 & Black & -0.741 & 0.133 & [-1.002, -0.480] \\
auth\_2 & t1 & Black & -0.803 & 0.135 & [-1.068, -0.538] \\
auth\_3 & t1 & Black & -0.439 & 0.124 & [-0.683, -0.195] \\
auth\_4 & t1 & Black &  0.058 & 0.120 & [-0.178, 0.293] \\

\hline
\multicolumn{6}{c}{\textbf{Anchor Items}} \\
\hline

anchor1m & t1 & White & -0.018 & 0.043 & [-0.102, 0.067] \\
anchor1m & t2 & White &  0.348 & 0.044 & [0.261, 0.434] \\

anchor1m & t1 & Black & -0.268 & 0.122 & [-0.506, -0.029] \\
anchor1m & t2 & Black &  0.292 & 0.122 & [0.052, 0.531] \\

anchor2m & t1 & White & -1.496 & 0.066 & [-1.625, -1.366] \\
anchor2m & t2 & White & -1.031 & 0.053 & [-1.134, -0.928] \\

anchor2m & t1 & Black & -1.451 & 0.180 & [-1.803, -1.099] \\
anchor2m & t2 & Black & -1.179 & 0.156 & [-1.484, -0.873] \\

anchor3m & t1 & White &  0.276 & 0.044 & [0.190, 0.362] \\
anchor3m & t2 & White &  0.953 & 0.051 & [0.853, 1.053] \\

anchor3m & t1 & Black &  0.173 & 0.121 & [-0.063, 0.410] \\
anchor3m & t2 & Black &  0.937 & 0.141 & [0.660, 1.215] \\

anchor4m & t1 & White & -1.924 & 0.089 & [-2.099, -1.749] \\
anchor4m & t2 & White & -1.208 & 0.057 & [-1.320, -1.097] \\

anchor4m & t1 & Black & -1.919 & 0.248 & [-2.404, -1.433] \\
anchor4m & t2 & Black & -1.179 & 0.156 & [-1.484, -0.873] \\

\hline
\end{tabular}
\end{table}

\begin{table}[H]
\centering
\caption{Threshold Differences (2008 CCAP)}
\label{tab:anes16_thresholds}
\small

\begin{tabular}{lcccc}
\hline
Item & t1 & t2 & t3 & Type \\
\hline

Item 1   & -0.524 / -0.741 & -- & -- & Child-rearing \\
Item 2   & -0.215 / -0.803 & -- & -- & Child-rearing \\
Item 3   &  0.164 / -0.439 & -- & -- & Child-rearing \\
Item 4   &  0.607 /  0.058 & -- & -- & Child-rearing \\
\hline

Anchor 1 & -0.018 / -0.268 &  0.348 / 0.292 & -- & Anchor \\
Anchor 2 & -1.496 / -1.451 & -1.031 / -1.179 & -- & Anchor \\
Anchor 3 &  0.276 /  0.173 &  0.953 / 0.937 & -- & Anchor \\
Anchor 4 & -1.924 / -1.919 & -1.208 / -1.179 & -- & Anchor \\
\hline

\end{tabular}
{\footnotesize
\par\noindent\raggedright
\textit{Note:} Values are reported as White / Black respondents.
}
\end{table}

\subsection*{2008 ANES}

\begin{table}[H]
\centering
\scriptsize
\caption{Estimated Thresholds (2008 ANES)}
\label{tab:thresholds_2008_anes_full}

\begin{tabular}{llcccc}
\hline
Item & Threshold & Group & Estimate & SE & 95\% CI \\
\hline

\multicolumn{6}{c}{\textbf{Child-Rearing Items}} \\
\hline

auth\_1 & t1 & White & -0.879 & 0.046 & [-0.970, -0.788] \\
auth\_2 & t1 & White & -0.554 & 0.043 & [-0.638, -0.471] \\
auth\_3 & t1 & White & -0.179 & 0.041 & [-0.259, -0.100] \\
auth\_4 & t1 & White &  0.216 & 0.041 & [0.137, 0.296] \\

auth\_1 & t1 & Black & -1.325 & 0.090 & [-1.501, -1.149] \\
auth\_2 & t1 & Black & -1.294 & 0.089 & [-1.467, -1.120] \\
auth\_3 & t1 & Black & -1.000 & 0.078 & [-1.152, -0.847] \\
auth\_4 & t1 & Black & -0.338 & 0.066 & [-0.467, -0.209] \\

\hline
\multicolumn{6}{c}{\textbf{Anchor Items}} \\
\hline

anchor1 & t1 & White & -0.831 & 0.046 & [-0.920, -0.741] \\
anchor1 & t2 & White &  0.179 & 0.041 & [0.100, 0.259] \\
anchor1 & t3 & White &  1.460 & 0.060 & [1.341, 1.578] \\

anchor1 & t1 & Black & -0.828 & 0.073 & [-0.972, -0.685] \\
anchor1 & t2 & Black & -0.086 & 0.065 & [-0.213, 0.040] \\
anchor1 & t3 & Black &  1.180 & 0.084 & [1.016, 1.344] \\

anchor2 & t1 & White & -1.117 & 0.051 & [-1.217, -1.017] \\
anchor2 & t2 & White &  0.616 & 0.043 & [0.531, 0.700] \\

anchor2 & t1 & Black & -1.000 & 0.078 & [-1.152, -1.000] \\
anchor2 & t2 & Black &  0.558 & 0.068 & [0.424, 0.692] \\

anchor3 & t1 & White & -1.048 & 0.049 & [-1.144, -0.951] \\
anchor3 & t2 & White & -0.072 & 0.040 & [-0.151, 0.007] \\
anchor3 & t3 & White &  0.926 & 0.047 & [0.833, 1.018] \\
anchor3 & t4 & White &  1.882 & 0.081 & [1.724, 2.040] \\

anchor3 & t1 & Black & -0.662 & 0.070 & [-0.799, -0.525] \\
anchor3 & t2 & Black &  0.160 & 0.065 & [0.033, 0.287] \\
anchor3 & t3 & Black &  0.866 & 0.074 & [0.721, 1.012] \\
anchor3 & t4 & Black &  1.617 & 0.107 & [1.408, 1.827] \\

anchor4 & t1 & White &  0.059 & 0.040 & [-0.020, 0.138] \\
anchor4 & t1 & Black &  0.241 & 0.065 & [0.113, 0.369] \\

\hline
\end{tabular}
\end{table}

\begin{table}[H]

\centering
\caption{Threshold Differences (2008 ANES)}
\label{tab:anes08_thresholds}
\small

\begin{tabular}{lcccc}
\hline
Item & t1 & t2 & t3 & Type \\
\hline

Item 1   & -0.879 / -1.325 & -- & -- & Child-rearing \\
Item 2   & -0.554 / -1.294 & -- & -- & Child-rearing \\
Item 3   & -0.179 / -1.000 & -- & -- & Child-rearing \\
Item 4   &  0.216 / -0.338 & -- & -- & Child-rearing \\
\hline

Anchor 1 & -0.831 / -0.828 &  0.179 / -0.086 &  1.460 / 1.180 & Anchor \\
Anchor 2 & -1.117 / -1.000 &  0.616 / 0.558 & -- & Anchor \\
Anchor 3 & -1.048 / -0.662 & -0.072 / 0.160 &  0.926 / 0.866 & Anchor \\
Anchor 4 &  0.059 /  0.241 & -- & -- & Anchor \\
\hline

\end{tabular}
{\footnotesize
\par\noindent\raggedright
\textit{Note:} Values are reported as White / Black respondents.
}
\end{table}

\subsection*{2016 ANES}

\begin{table}[H]
\centering
\scriptsize
\caption{Estimated Thresholds (2016 ANES)}
\label{tab:thresholds_2016_anes}

\begin{tabular}{llcccc}
\hline
Item & Threshold & Group & Estimate & SE & 95\% CI \\
\hline

\multicolumn{6}{c}{\textbf{Child-Rearing Items}} \\
\hline

auth\_1 & t1 & White & -0.524 & 0.045 & [-0.613, -0.435] \\
auth\_2 & t1 & White & -0.215 & 0.043 & [-0.300, -0.130] \\
auth\_3 & t1 & White &  0.164 & 0.043 & [ 0.079,  0.249] \\
auth\_4 & t1 & White &  0.607 & 0.046 & [ 0.517,  0.698] \\

auth\_1 & t1 & Black & -0.741 & 0.133 & [-1.002, -0.480] \\
auth\_2 & t1 & Black & -0.803 & 0.135 & [-1.068, -0.538] \\
auth\_3 & t1 & Black & -0.439 & 0.124 & [-0.683, -0.195] \\
auth\_4 & t1 & Black &  0.058 & 0.120 & [-0.178,  0.293] \\

\hline
\multicolumn{6}{c}{\textbf{Anchor Items}} \\
\hline

anchor1m & t1 & White & -0.018 & 0.043 & [-0.102,  0.067] \\
anchor1m & t2 & White &  0.348 & 0.044 & [ 0.261,  0.434] \\

anchor2m & t1 & White & -1.496 & 0.066 & [-1.625, -1.366] \\
anchor2m & t2 & White & -1.031 & 0.053 & [-1.134, -0.928] \\

anchor3m & t1 & White &  0.276 & 0.044 & [ 0.190,  0.362] \\
anchor3m & t2 & White &  0.953 & 0.051 & [ 0.853,  1.053] \\

anchor4m & t1 & White & -1.924 & 0.089 & [-2.099, -1.749] \\
anchor4m & t2 & White & -1.208 & 0.057 & [-1.320, -1.097] \\

\hline

auth\_1 & t1 & Black & -0.741 & 0.133 & [-1.002, -0.480] \\
auth\_2 & t1 & Black & -0.803 & 0.135 & [-1.068, -0.538] \\
auth\_3 & t1 & Black & -0.439 & 0.124 & [-0.683, -0.195] \\
auth\_4 & t1 & Black &  0.058 & 0.120 & [-0.178,  0.293] \\

anchor1m & t1 & Black & -0.268 & 0.122 & [-0.506, -0.029] \\
anchor1m & t2 & Black &  0.292 & 0.122 & [ 0.052,  0.531] \\

anchor2m & t1 & Black & -1.451 & 0.180 & [-1.803, -1.099] \\
anchor2m & t2 & Black & -1.179 & 0.156 & [-1.484, -0.873] \\

anchor3m & t1 & Black &  0.173 & 0.121 & [-0.063,  0.410] \\
anchor3m & t2 & Black &  0.937 & 0.141 & [ 0.660,  1.215] \\

anchor4m & t1 & Black & -1.919 & 0.248 & [-2.404, -1.433] \\
anchor4m & t2 & Black & -1.179 & 0.156 & [-1.484, -0.873] \\

\hline
\end{tabular}
\end{table}

\begin{table}[H]
\centering
\caption{Threshold Differences (2016 ANES)}
\label{tab:anes08_thresholds}
\small

\begin{tabular}{lcccc}
\hline
Item & t1 & t2 & t3 & Type \\
\hline

Item 1   & -0.524 / -0.741 & -- & -- & Child-rearing \\
Item 2   & -0.215 / -0.803 & -- & -- & Child-rearing \\
Item 3   &  0.164 / -0.439 & -- & -- & Child-rearing \\
Item 4   &  0.607 /  0.058 & -- & -- & Child-rearing \\
\hline

Anchor 1 & -0.831 / -0.828 &  0.179 / -0.086 &  1.460 / 1.180 & Anchor \\
Anchor 2  & -1.117 / -1.000 &  0.616 / 0.558 & -- & Anchor \\
Anchor 3 & -1.048 / -0.662 & -0.072 / 0.160 &  0.926 / 0.866 & Anchor \\
Anchor 4 &  0.059 /  0.241 & -- & -- & Anchor \\
\hline

\end{tabular}
{\footnotesize
\par\noindent\raggedright
\textit{Note:} Values are reported as White / Black respondents.
}
\end{table}


\begin{thebibliography}{}

\bibitem[Byrne, Shavelson, and Muthén(1989)]{Muthen1989}
Byrne, B. M., R. J. Shavelson, and B. O. Muthén. 1989. 
“Testing for the Equivalence of Factor Covariance and Mean Structures: The Issue of Partial Measurement Invariance.” 
\textit{Psychological Bulletin} 105(3): 456--466.

\bibitem[Dawson(1994)]{dawson1994}
Dawson, Michael C. 1994. 
\textit{Behind the Mule: Race and Class in African-American Politics}. 
Princeton: Princeton University Press.

\bibitem[Davidov et al.(2014)]{Davidov2014}
Davidov, E., B. Meuleman, J. Cieciuch, P. Schmidt, and J. Billiet. 2014. 
“Measurement Equivalence in Cross-National Research.” 
\textit{Annual Review of Sociology} 40: 55--75.

\bibitem[Dimitrov(2006)]{dimitrov2006}
Dimitrov, D. M. 2006. 
“Comparing Groups on Latent Variables: A Structural Equation Modeling Approach.” 
\textit{Work} 26: 429--436.

\bibitem[Engelhardt, Feldman, and Hetherington(2021)]{engelhardt2021}
Engelhardt, A. M., S. Feldman, and M. J. Hetherington. 2021. 
“Advancing the Measurement of Authoritarianism.” 
\textit{Political Behavior} 45: 537--560.

\bibitem[Feldman(2003)]{RN1066}
Feldman, S. 2003. 
“Enforcing Social Conformity: A Theory of Authoritarianism.” 
\textit{Political Psychology} 24(1): 41--74.

\bibitem[Federico, Feldman, and Weber(2026)]{Federico2026}
Federico, Christopher M., Stanley Feldman, and Christopher Weber. 2026. 
\textit{The Authoritarian Divide: Partisan Identity, Voting, and the Transformation of the American Electorate}. 
New York: Oxford University Press. https://doi.org/10.1093/9780197813379.001.0001.

\bibitem[Feldman and Stenner(1997)]{Feldman1997}
Feldman, S., and K. Stenner. 1997. 
“Perceived Threat and Authoritarianism.” 
\textit{Political Psychology} 18: 741--770.

\bibitem[Hetherington and Suhay(2011)]{Hethering2011}
Hetherington, M. J., and E. Suhay. 2011. 
“Authoritarianism, Threat, and Americans' Support for the War on Terror.” 
\textit{American Journal of Political Science} 55(3): 546--560.

\bibitem[Hetherington and Weiler(2009)]{RN158}
Hetherington, M. J., and J. D. Weiler. 2009. 
\textit{Authoritarianism and Polarization in American Politics}. New York: Cambridge University Press.

\bibitem[Kim and Yoon(2011)]{KimYoon2011}
Kim, E. S., and M. Yoon. 2011. 
“Testing Measurement Invariance: A Comparison of Multiple-Group Categorical CFA and IRT.” 
\textit{Structural Equation Modeling} 18(2): 212--228.

\bibitem[Millsap(2011)]{RN1030}
Millsap, R. E. 2011. 
\textit{Statistical Approaches to Measurement Invariance}. New York: Routledge.

\bibitem[Mund, Johnson, and Nestler(2021)]{RN1109}
Mund, Michael, Matthew D. Johnson, and Steffen Nestler. 2021. ``Changes in Size and Interpretation of Parameter Estimates in Within-Person Models in the Presence of Time-Invariance and Time-Varying Covariate.'' \textit{Frontiers in Psychology}. https://doi.org/10.3389/fpsyg.2021.666928.

\bibitem[Muthén(1984)]{Muthen1984}
Muthén, B. O. 1984. 
“A General Structural Equation Model with Dichotomous, Ordered Categorical, and Continuous Latent Variable Indicators.” 
\textit{Psychometrika} 49(1): 115--132.

\bibitem[Muthén(2012)]{Muthen2012}
Muthén, B. O. 2012. 
“Advances in Structural Equation Modeling with Categorical Variables.” 
In \textit{Mplus Short Courses}.

\bibitem[Muthén and Muthén(2007)]{Muthen2007}
Muthén, L. K., and B. O. Muthén. 2007. 
\textit{Mplus User’s Guide}. 5th ed. Los Angeles, CA: Muthén \& Muthén.

\bibitem[Pérez and Hetherington(2014)]{RN925}
Pérez, E. O., and M. J. Hetherington. 2014. 
“Authoritarianism in Black and White: Testing the Cross-Racial Validity of the Child Rearing Scale.” 
\textit{Political Analysis} 22(3): 398--412.

\bibitem[Pietryka and MacIntosh(2022)]{Pietryka2022}
Pietryka, M. T., and R. C. MacIntosh. 2022. 
“ANES Scales Often Do Not Measure What You Think They Measure.” 
\textit{Journal of Politics} 84(2): 1074--1090.

\bibitem[Rosseel(2012)]{rosseel2012}
Rosseel, Y. 2012. 
“lavaan: An R Package for Structural Equation Modeling.” 
\textit{Journal of Statistical Software} 48(2): 1--36.

\bibitem[Steenkamp and Baumgartner(1998)]{Steenkamp1998}
Steenkamp, J.-B. E. M., and H. Baumgartner. 1998. 
“Assessing Measurement Invariance in Cross-National Consumer Research.” 
\textit{Journal of Consumer Research} 25(1): 78--90.

\bibitem[King et al.(2004)]{King2004}
King, Gary, Christopher J. L. Murray, Joshua A. Salomon, and Ajay Tandon. 2004. 
“Enhancing the Validity and Cross-Cultural Comparability of Measurement in Survey Research.” 
\textit{American Political Science Review} 98(1): 191--207.

\bibitem[Stenner(2005)]{Stenner2005}
Stenner, K. 2005. 
\textit{The Authoritarian Dynamic}. Cambridge: Cambridge University Press.

\bibitem[Vandenberg and Lance(2000)]{Vandenberg2000}
Vandenberg, R. J., and C. E. Lance. 2000. 
“A Review and Synthesis of the Measurement Invariance Literature.” 
\textit{Organizational Research Methods} 3(1): 4--70.

\bibitem[Jefferson(2023)]{Jefferson2023}
Jefferson, H. 2023. 
“The Politics of Respectability and Black Americans’ Punitive Attitudes.” 
\textit{American Political Science Review} 117(4): 1448--1464.

\bibitem[Harris(2012)]{Harris2012}
Harris, F. C. 2012. 
\textit{The Price of the Ticket: Barack Obama and the Rise and Decline of Black Politics}. 
Oxford University Press.

\bibitem[Cohen(1999)]{Cohen1999}
Cohen, C. J. 1999. 
\textit{The Boundaries of Blackness: AIDS and the Breakdown of Black Politics}. 
University of Chicago Press.

\bibitem[White and Laird(2020)]{WhiteLaird2020}
White, I. K., and C. N. Laird. 2020. 
\textit{Steadfast Democrats: How Social Forces Shape Black Political Behavior}. 
Princeton University Press.

\bibitem[Krosnick(1991)]{Krosnick1991}
Krosnick, Jon A. 1991. 
“Response Strategies for Coping with the Cognitive Demands of Attitude Measures in Surveys.” 
\textit{Applied Cognitive Psychology} 5(3): 213--236.

\bibitem[Zaller(1992)]{Zaller1992}
Zaller, John R. 1992. 
\textit{The Nature and Origins of Mass Opinion}. 
Cambridge: Cambridge University Press.


\bibitem[Parker and Towler(2019)]{ParkerTowler2019}
Parker, Christopher Sebastian, and Christopher C. Towler. 2019. 
\textit{Race and Authoritarianism in American Politics}. 
Annual Review of Political Science 22:503--519. 
https://doi.org/10.1146/annurev-polisci-050317-064519

\bibitem[Zaller and Feldman(1992)]{ZallerFeldman1992}
Zaller, John, and Stanley Feldman. 1992. 
“A Simple Theory of the Survey Response: Answering Questions versus Revealing Preferences.” 
\textit{American Journal of Political Science} 36(3): 579--616.


\end{thebibliography}

\newpage

\begin{thebibliography}{9}


\bibitem[Millsap and Kwok(2004)]{Millsap2004}
Millsap, R. E., and O. M. Kwok. 2004. 
"Evaluating the Impact of Partial Measurement Invariance on Latent Mean Differences." 
\textit{Structural Equation Modeling} 11: 515--534.

\bibitem[Byrne(1989)]{Byrne1989}
Byrne, B. M. 1989. 
\textit{A Primer of LISREL}. Springer.


\bibitem[Davidov et al.(2014)]{Davidov2014}
Davidov, E., B. Meuleman, J. Cieciuch, P. Schmidt, and J. Billiet. 2014. 
“Measurement Equivalence in Cross-National Research.” 
\textit{Annual Review of Sociology} 40: 55--75.

\bibitem[Steenkamp and Baumgartner(1998)]{Steenkamp1998}
Steenkamp, J.-B. E. M., and H. Baumgartner. 1998. 
“Assessing Measurement Invariance in Cross-National Consumer Research.” 
\textit{Journal of Consumer Research} 25(1): 78--90.

\bibitem[Vandenberg and Lance(2000)]{Vandenberg2000}
Vandenberg, R. J., and C. E. Lance. 2000. 
“A Review and Synthesis of the Measurement Invariance Literature.” 
\textit{Organizational Research Methods} 3(1): 4--70.

\end{thebibliography}
\end{document}